\documentclass[a4paper,11pt,oneside,onecolumn]{article}
\usepackage{amsmath,amsthm,amsfonts,latexsym,amscd,amssymb, graphicx, color,enumerate, caption, longtable, booktabs, array}
\usepackage{xcolor}

\DeclareUnicodeCharacter{200B}{}
\usepackage[hidelinks]{hyperref}
\newtheorem{thm}{Theorem}[section]
\newtheorem{dfn}[thm]{Definition}
\newtheorem{rmk}[thm]{Remark}
\newtheorem{exl}[thm]{Example}

\newtheorem{propo}[thm]{Proposition}
\newtheorem{coro}[thm]{Corollary}

\newtheorem{lem}[thm]{Lemma}
\newtheorem{qn}[thm]{Question}

\begin{document}
	\title{\bf {Constructions of reversible linear and DNA codes through $T_{D,g}^n-$set and $(n_1,c_1|n_2,c_2)-$right polycyclic codes}}  
	\author{\bf Avanish Kumar Chaturvedi$^{a}$\footnote{Corresponding author} and \bf Satyadeep Pandey$^{b}$} 
	\date{Department of Mathematics, University of Allahabad \\ Prayagraj-211002, India \\ {\footnotesize  $^{a}$akchaturvedi.math@gmail.com, achaturvedi@allduniv.ac.in\\
			$^{b}$ pandeysatyam14082000@gmail.com}}     	
	\maketitle		
	\begin{abstract}
		In this paper we give a construction technique of $T_{D,g}^n-$set for reversible linear codes and reversible DNA codes with their code parameters. This technique yields many reversible MDS codes which significantly improves the $T^n-$set construction by Oztas et.al. in terms of obtained code parameters. Concatenation-based constructions are presented for reversible linear codes satisfying self-duality and linear complementary dual (LCD) conditions, giving necessary and sufficient condition for desired dual properties. 
		
		We introduce a new notion of $(n_1,c_1|n_2,c_2)-$ right polycyclic codes. Also, a new construction technique is developed to obtain reversible linear $(n_1,c_1|n_2,c_2)-$ right polycyclic codes with good parameters (MDS codes) over any finite commutative rings.
	\end{abstract}
	
	\section{Introduction}
	DNA (deoxyribonucleic acid) carries the genetic information of living organisms and consists of sequences formed from four nucleotides: adenine (A), cytosine (C), guanine (G), and thymine (T). The complementary pairing of these nucleotides follows the Watson--Crick rule, according to which A pairs with T and C pairs with G.
	The feasibility of DNA-based computation was initially established by Adleman \cite{Adleman}, who employed DNA strands to solve the combinatorial optimization problem known as the traveling salesman problem. Subsequently, DNA-based approaches have been investigated for various applications, including cryptographic analysis of the Data Encryption Standard (DES) \cite{AdlemanDES}, \cite{Boneh} and data storage \cite{Mansuripur}. 
	
	Nevertheless, the practical implementation of DNA-based systems necessitates the careful design of DNA sequences to prevent undesirable hybridization and the formation of secondary structures \cite{Mansuripur}. This has motivated extensive research on the construction of DNA codes that satisfy multiple constraints, including minimum Hamming distance, reversibility, complementarity, and balanced nucleotide composition, commonly expressed through the GC-content constraint. Several construction techniques have been proposed to develop DNA codes that address different combinations of these requirements \cite{Abolution}, \cite{Frutos}, \cite{Gaborit}, \cite{King}, \cite{Smith} and \cite{Song}.
	
	In \cite{Oztas} authors use the concept of double reflected matrices and concatenation  of generator matrices to obtain reversible linear and reversible DNA codes. In \cite{Chaturvedi}, the concept of $D-$ reflected matrices is introduced which significantly extends the scope of construction of reversible linear and DNA codes and provides improved code parameters including various reversible linear MDS codes. In \cite{Oztas2}, $T^n$ set construction is used to generate reversible linear and reversible DNA codes.\\
	
	Motivated by all these developments, we give a generalized construction technique of $T_{D,g}^n-$set (see Definition \ref{generalized set}) for reversible linear and DNA codes, which is a generalization of the $T^n$-set construction given in \cite{Oztas} for reversible linear codes. The new generalized technique employs an arbitrary involutory matrix $D$ and hence broadens the scope in comparision to the $T^n-$set construction of \cite{Oztas}. In support, see Example \ref{better linear parameter}, in which using Sage Math computation, MDS code is obtained using $T_{D,g}^n$-set for an appropriate $D$ while through $T^n-$ set  yields an AMDS code. Also, in Remark \ref{paper 3 mds amds number}, using SageMath software \cite{Sage}, we obtain 256 generator matrices over $F_{16}$ for a given polynomial. Out of $256$ reversible codes obtained $181$ are AMDS and $75$ are MDS. 	 Thus, the $T_{D,g}^n-$set construction gives better code parameters (MDS codes) preserving reversibility.     Many results (Corollary \ref{linear corollary}) regarding the $T^n-$set construction in \cite{Oztas2} are shown to be a particular case of the $T_{D,g}^n-$set construction (see Theorem \ref{linear theorem}).

	Also, using Hadamard product of matrices the $T_{D,g}^n$-set construction is extended to reversible DNA codes (see Theorem \ref{DNA theorem}). In Example \ref{DNA Example}, a reversible DNA code is obtained which is also MDS while the $T^n-$set construction yields an AMDS reversible DNA code.
	
	The dual code of a linear code provides important information about its algebraic structure and its relationship with the ambient vector space. The dual $C^\perp$ of a linear code $C$ is the linear space of all the vectors which are orthogonal to the given code with respect to the usual product. The relationship between a code and its dual gives rise to several important classes of codes. A code is self-orthogonal if $C\subseteq C^\perp$, whereas it is self-dual if $C=C^\perp$. A code satisfying $C^\perp\subseteq C$ is called dual-containing. If $C\cap C^\perp=\{0\},$ then $C$ is called linear complementary dual (LCD) code \cite{Massey}. These relationships are useful in studying the structure of linear codes and in constructing codes with additional algebraic constraints.
		
	In the present work, we extend the $T_{D,g}^n-$set construction to obtain reversible code which are also self dual code. Precisely, we give a necessary and sufficient condition (see Theorem \ref{reversible self dual}) on a polynomial $g$ so that the $T_{D,g}^n-$ set construction gives reversible self dual code. Moreover, we give a characterization for the $D-$reflected matrix construction of \cite{Chaturvedi}, to obtain reversible LCD linear code (see Theorem \ref{lcd characterisation}). \\

	The notion of polycylic codes is classic and it is a generalization of cyclic codes. In recent years, polycyclic codes have attracted increasing research interest, as illustrated by several studies in the literature \cite{Aydin, Fotue, Shi M., Shi, Wu}. Also, various generalizations of polycyclic codes have been studied. In \cite[Definition 4]{Suxsena}, Suxena et. al. introduced the concept of right generalized quasi polycyclic codes over a finite field $F_q.$ They use independent polycylic shifts in different blocks. Further, in \cite{Saif}, the notion of generalized quasi polycyclic codes is further generalized over a specific finite ring $F_q+uF_q$ with $u^2=0.$
	
	 In the present work, we introduce a new notion of $(n_1,c_1|n_2,c_2)-$right polycyclic codes over any arbitrary finite commutative ring with unity (see Definition \ref{(n1,c1)}).  We show that an $(n_1,c_1|n_2,c_2)-$right polycyclic code need not be polycyclic (see Example \ref{partwise poly need not be poly}) . Moreover, if two generator matrices separately generate polycyclic codes then it is not necessary that their concatenation generates $(n_1,c_1|n_2,c_2)-$right polycyclic code (see Example \ref{concatenation of poly not partwise poly}). We provide a very fruitful sufficient condition for the concatenation of two matrices to generate $(n_1,c_1|n_2,c_2)-$right polycyclic code given each of them individually generates a polycyclic code (see Proposition \ref{5}). We raise a natural open Question \ref{open} that "Is the converse of Proposition \ref{5} true?"
	 Further, we give a construction technique for $(n_1,c_1|n_2,c_2)-$right polycyclic codes which are also reversible (see Theorem \ref{poly main construction}). Also, several MDS codes are obtained using SageMath computations (see Example \ref{main example poly}).\\

	Thus the proposed approaches provide a broader range of constructions and can yield improved code parameters, including MDS codes in suitable cases. Further,  These constructions illustrate different ways of combining reversibility with duality conditions, and provide a basis for further investigation of code parameters and potential applications in DNA-based storage, error correction, and code-based cryptography.	
	
	\section{Preliminaries}
	In this section, we recall some basic definitions and notations relevant to the present work. 	
	Let $R$ be a finite commutative ring with identity. Following \cite{Oztas}, an \emph{$R$-linear code} $C$ of length $n$ is an $R$-submodule $C \subseteq R^n$ and for a vector $c = (c_o,c_1,\ldots,c_{n-1}) \in R^n$, the \emph{reverse} of $c$ is defined as $ c^r = (c_{n-1},c_{n-2},\ldots,c_0).$
	\begin{dfn}\cite[Definition 1]{Oztas}
		A linear code $C$ of length $n$ over $R$ is said to be a \emph{reversible} code if for any $c \in C,\:c^r \in C.$
	\end{dfn}
	
	\begin{dfn}\cite[Definition 2]{Oztas}\label{double reflection definition} Let $M = [a_{ij}]_{m \times n}$ be an $m \times n$ matrix over the ring $R$. Then, the \emph{double-reflected matrix} of $M$ is denoted by $M''$ and defined as $
		M'' = [b_{ij}]_{m \times n},$
		where $b_{ij} = a_{(m-i+1)(n-j+1)},$
		that is,
		$
		M'' =
		\begin{bmatrix}
			a_{mn} & \cdots & a_{m3} & a_{m2} & a_{m1} \\
			\vdots &        & \vdots & \vdots & \vdots \\
			\vdots &        & \vdots & \vdots & \vdots \\
			a_{2n} & \cdots & a_{23} & a_{22} & a_{21} \\
			a_{1n} & \cdots & a_{13} & a_{12} & a_{11}
		\end{bmatrix}_{m \times n}.
		$
	\end{dfn}
	Following \cite{Brawley}, a matrix $A \in (R)_n,$ is called \emph{involutory} iff $A^2= I_n,$ where $I_n$ is the $n\times n$	identity matrix, $R$ is a finite commutative ring with identity and $(R)_n$ the set of $n\times n$ matrices over $R.$
	\begin{dfn}\cite[Definition 3]{Oztas}: Let $M = [a_{ij}]_{m\times n}$ and 
		$N = [b_{ij}]_{m\times n}$ be two $m \times n$ matrices over the ring $R$. 
		Then, the \emph{Hadamard product} of $M$ and $N$ is denoted by $M \circ N$ and is defined by
		\[
		M \circ N = [a_{ij}b_{ij}]_{m\times n},
		\]
		where $a_{ij}, b_{ij} \in R$ with $1 \le i \le m$, $1 \le j \le n$.
	\end{dfn}
	Following \cite{Oztas}, we have some properties of Hadmard product.
	
	(i) The Hadamard $s$-power of $M = [a_{ij}]_{m\times n}$ is denoted by 
	$M^{\circ s}$, and defined as
	$
	M^{\circ s} = [(a_{ij})^s]_{m\times n}.
	$
	
	(ii) Since $R$ is a finite commutative ring, so	$M \circ N = N \circ M.$
	\begin{dfn}\label{concatenation}\cite[Definition 1]{Chaturvedi}
		Let $A$ and $B$ be two $k\times n$ matrices over a ring $R$. By \emph{concatenation} of $A$ and $B$ we mean a row wise concatenation $[A|B]$, of order $k\times 2n.$
	\end{dfn}
	\begin{lem}\cite[Lemma 2.5]{Chaturvedi}\label{product of matrices}
		Let $R$ be a finite commutative ring with unity. For positive integers $k$ and $n$, let $J_n$ be the $n \times n$ permutation matrix with ones on the anti-diagonal and zeros elsewhere and $M$ be a $k\times n$ matrix over $R.$ Then
		\begin{enumerate}[(1)]
			\item $J_n^2$ is the $n\times n$ identity matrix over $R$, so, $J_n$ is an involutory matrix over $R$.
			\item For any $c=(c_0,\: c_1,...\:,c_{n-1})\in R^n$, the reverse of $c$ is $c^r=cJ_n$, where $c^r=(c_{n-1},\: c_{n-2},\:...,c_0)$.
			\item $J_kM=M_1$, where $M_1$ is the matrix obtained by reversing the order of the rows of $M$.
			\item $MJ_n=M_2$, where $M_2$ is the matrix obtained by reversing order of the columns of $M$.
			\item $J_k M J_n=M'',$ where $M''$ is obtained by reversing the order of rows and columns of $M$ simultaneously, is the double reflected matrix of $M$.
			\item $[A\mid B]J_{2n}=[BJ_n\mid AJ_n]$ for any $k\times n$ matrices $A$ and $B$.
			\item $A^{os}J_n=(AJ_n)^{os}$, for any $k\times n$ matrix $A$, where $A^{os}$ denotes the Hadamard $s$ power of the matrix $A$.
		\end{enumerate}
	\end{lem}
	\begin{dfn}\cite[Definition 3.1]{Chaturvedi}\label{reflected}\label{D-reflection definition}
		Let $M$ be a $k\times n$ matrix over a ring $R$ and $D$ be any involutory $k\times k$ matrix over $R$, that is, $D\in R^{k\times k}$ such that $D^2=I_d.$ The $D-$ reflected matrix of $M$ denoted by $M'_D$ is defined as the matrix product $DMJ_n$, where $J_n$ is the permutation matrix of order $n$ over $R$ with ones as anti diagonal entries and zeros elsewhere. 
	\end{dfn}
	
	\begin{dfn}\cite[Definition 5.4.3]{Xing}
		A linear code with parameters $[n,k,d]$ such that $k+d = n+1$ 
		is called a maximum distance separable (MDS) code.
	\end{dfn}
	\begin{dfn}\cite[Definition 1]{De Broer} The Singleton defect of an $[n,k,d]$ code $C$ is $s(C)=n-k+1-d.$
	\end{dfn}
	\begin{dfn}\cite[Definition 5]{De Broer} A code with Singleton defect $s(C)=1$ is almost MDS (AMDS for short).
	\end{dfn}
	
	Recall from \cite[Definition 1]{Oztas2}, let $R$ be a finite commutative ring with unity and let $g(x)=z_0+z_1x+\cdots+z_sx^s\in R[x]$ be a polynomial with $\deg(g)=s<n$. Then the $T_g^n$ generator set with respect to $g(x)$ is defined as $T_g^n=\{g(x),xg(x),\ldots,x^{n-s-1}g(x)\}$.
	Its coefficient matrix is denoted by $G_{T_{g}^n}$ and is given by $$
	G_{T_{g}^n}=
	\begin{bmatrix}
		z_0&z_1&\cdots&z_s&0&\cdots&0\\
		0&z_0&z_1&\cdots&z_s&\cdots&0\\
		\vdots&&\ddots&&&&\vdots\\
		0&\cdots&0&z_0&z_1&\cdots&z_s
	\end{bmatrix}.$$
	Let $g^*(x)=z_s+z_{s-1}x+\cdots+z_0x^s$ denote the reciprocal polynomial of $g(x)$. Then the $T_{g^*}^n$ generator set with respect to $g^*(x)$ is $\{g^*(x),xg^*(x),\ldots,x^{n-s-1}g^*(x)\}$. The matrix form of $T_{g^*}^n$ generator set is $$G_{{T_{g^*}^n}} = 
	\left[
	\begin{array}{c}
		g^*(x) \\
		x g^*(x) \\
		x^2 g^*(x) \\
		\vdots \\
		x^{n-s-2} g^*(x) \\
		x^{n-s-1} g^*(x)
	\end{array}
	\right]$$$$
	=
	\left[
	\begin{array}{ccccccccccc}
		z_s & z_{s-1} & z_{s-2} & \dots & z_0 & 0 & 0 & \dots & 0 & 0 & 0 \\
		0 & z_s & z_{s-1} & z_{s-2} & \dots & z_0 & 0 & \dots & 0 & 0 & 0 \\
		0 & 0 & z_s & z_{s-1} & z_{s-2} & \dots & z_0 & \dots & 0 & 0 & 0 \\
		\dots & \dots & \dots & \dots & \dots & \dots & \dots & \dots & \dots & \dots & \dots \\
		0 & 0 & \dots & z_s & z_{s-1} & z_{s-2} & \dots & z_2 & z_1 & z_0 & 0 \\
		0 & 0 & \dots & 0 & z_s & z_{s-1} & z_{s-2} & \dots & z_2 & z_1 & z_0
	\end{array}
	\right].$$
	
	\section{$G_{T_{D,g}^n}$-set construction for reversible linear and reversible DNA codes}
	
	Let $J_m$ and $J_n$ denote the reversal matrices of orders $m$ and $n$, respectively as in Lemma \ref{product of matrices}. Then by Lemma \ref{product of matrices} (5), we observe that the coefficient matrices associated with $g(x)$ and $g^*(x)$ satisfy the relation $G_{T_{g^*}^n}=J_mG_{T_g^n}J_n.$
	Thus, the second block in the generator matrix of the construction in \cite{Oztas2} is precisely the double reflection of $G_{T_{g}^n}$. The concatenation technique using double reflected matrices of \cite{Oztas}, has been generalized in \cite{Chaturvedi}, where $D-$ reflected matrices are used. Hence, the $T_g^n$ generator set admits a natural generalization as $T_{D,g}^n-$set, which we discuss below. First we describe the notion of $n-$length reciprocal of a polynomial, whose degree may be less than $n-1.$ 
	\begin{dfn}\label{length n reciprocal}
		Let $R$ be a commutative ring with unity and
		$f(x)=\sum_{i=0}^{n-1}a_i x^i\in R[x]$ be a polynomial
		of degree less than $n$. The \emph{length-$n$ reciprocal}
		of $f(x)$ is defined by
		$f^{*_{n}}(x)=x^{n-1}f(x^{-1})
		=\sum_{i=0}^{n-1}a_i x^{n-1-i}.$
	\end{dfn}	
	\begin{dfn}\label{generalized set}
		Let $R$ be a finite commutative ring with unity and 
		$g(x)=z_0+z_1x+\cdots+z_sx^s\in R[x]$ be a polynomial with $\deg(g)=s<n$. Let $m=n-s$.	Let $D\in R^{m\times m}$. The $D$-generalized $T^n$-set associated with $g(x)$ is the family denoted by	$T_{D,g}^n=\{f_0^{*_{n}}(x),f_1^{*_{n}}(x),\ldots,f_{m-1}^{*_{n}}(x)\},$ where $f_i(x)=\sum_{j=0}^{m-1}d_{ij}x^jg(x),$ for each $i=0,1,...,m-1.$
	\end{dfn}
	\begin{dfn}\label{coefficient matrix}
		The coefficient matrix $G_{T_{D,g}^n}$ of $T_{D,g}^n$ is the matrix whose rows are the coefficients of the polynomials in the family $T_{D,g}^n$ in the given order.
	\end{dfn}
	\begin{lem}\label{matrix row}
		Let $R$ be a finite commutative ring with unity,
		$A=(a_{ij})\in R^{m\times n}$ and
		$B=(b_{ij})\in R^{n\times p}$. Then, for every
		$i\in\{1,\ldots,m\}$, the $i^{th}$ row of $AB$ is a
		linear combination of the rows of $B$, with coefficients
		given by the entries of the $i^{th}$ row of $A$.
	\end{lem}
	
	\begin{proof}
		Let $R_j(B)=(b_{j1},b_{j2},\ldots,b_{jp})$ denote the
		$j^{th}$ row of $B$. By the definition of matrix multiplication,
		the $(i,k)^{th}$ entry of $AB$ is
		$(AB)_{ik}=\sum_{j=1}^{n}a_{ij}b_{jk}, 1\leq k\leq p.$
		Consequently, the $i$-th row of $AB$ is
		\begin{align*}
			R_i(AB)
			&=\left(\sum_{j=1}^{n}a_{ij}b_{j1},
			\ldots,\sum_{j=1}^{n}a_{ij}b_{jp}\right)\\
			&=\sum_{j=1}^{n}a_{ij}
			(b_{j1},b_{j2},\ldots,b_{jp})\\
			&=\sum_{j=1}^{n}a_{ij}R_j(B)\\
			&=a_{i1}R_1(B)+\cdots+a_{in}R_n(B).
		\end{align*}
		Thus, the $i^{th}$ row of $AB$ is a linear combination of the
		rows of $B$, with coefficients given by the entries of the
		$i^{th}$ row of $A$.
	\end{proof}
	
	\begin{lem}\label{MainLemma}
		Let $R$ be a finite commutative ring with unity and $D\in R^{m\times m}$ be an involutory matrix. Then the coefficient matrix $G_{T_{D,g}^n}$ of $T_{D,g}^n$  is the $D-$ reflected matrix of $G_{T_{g}^n}$, that is, $G_{T_{D,g}^n}=DG_{T_g^n}J_n.$
		
	\end{lem}
	\begin{proof}
		We prove it  explicitly first by computing entries of the
		matrix $DG_{T_g^n}J_n$ and then showing that its $i^{th}$ row is precisely the coefficient vector of $f_i^{*_{n}}(x)$.		
		For $j=0,1,\ldots,m-1$, let $R_j$ be denoted by the $(j+1)^{th}$ row of
		$G_{T_g^n}$. Since the $(j+1)^{th}$polynomial in $T_g^n$ is $x^jg(x)$, we have
		$R_j=\operatorname{coeff}\bigl(x^jg(x)\bigr).$ More explicitly, $R_0=
		\begin{bmatrix}
			z_0&z_1&\cdots&z_s&0&\cdots&0
		\end{bmatrix},$ 
		$R_1=\begin{bmatrix}
			0&z_0&z_1&\cdots&z_s&0&\cdots&0
		\end{bmatrix}.$ Hence
		$G_{T_g^n}=
		\begin{bmatrix}
			R_0\\
			R_1\\
			\vdots\\
			R_{m-1}
		\end{bmatrix}.$
		Writing $D=
		\begin{bmatrix}
			d_{00}&d_{01}&\cdots&d_{0 (m-1)}\\
			d_{10}&d_{11}&\cdots&d_{1(m-1)}\\
			\vdots&\vdots&\ddots&\vdots\\
			d_{(m-1)0}&d_{(m-1)1}&\cdots&d_{(m-1)(m-1)}
		\end{bmatrix}.$
		Then by Lemma \ref{matrix row}, the $i^{th}$ row of $DG_{T_g^n}$ is $\operatorname{Row}_i(DG_{T_g^n})= d_{i0}R_0+d_{i1}R_1+\cdots+d_{i,m-1}R_{m-1}.$
		Since $R_j$ is the coefficient vector of $x^jg(x)$, the polynomial
		corresponding to the $i^{th}$ row of $DG_{T_g^n}$ is $d_{i0}g(x)+d_{i1}xg(x)+\cdots+d_{i,m-1}x^{m-1}g(x).$
		By definition of $f_i(x)$, this is $f_i(x)=\sum_{j=0}^{m-1}d_{ij}x^jg(x).$
		Therefore, if $f_i(x)=a_{i0}+a_{i1}x+\cdots+a_{i,n-1}x^{n-1},$
		then the $i$-th row of $DG_{T_g^n}$ is $\operatorname{Row}_i(DG_{T_g^n})=
		\begin{bmatrix}
			a_{i0}&a_{i1}&\cdots&a_{i(n-1)}
		\end{bmatrix}.$
		Consequently, $DG_{T_g^n}=
		\begin{bmatrix}
			a_{00}&a_{01}&\cdots&a_{0(n-1)}\\
			a_{10}&a_{11}&\cdots&a_{1(n-1)}\\
			\vdots&\vdots&\ddots&\vdots\\
			a_{(m-1)0}&a_{(m-1)1}&\cdots&a_{(m-1)(n-1)}
		\end{bmatrix}.$
		By Lemma \ref{product of matrices} (2), multiplication of a row vector $(a_0,a_1,\ldots,a_{n-1})$ by $J_n$ gives $(a_0,a_1,\ldots,a_{n-1})J_n=(a_{n-1},a_{n-2},\ldots,a_1,a_0).$
		Therefore, $$DG_{T_g^n}J_n=
		\begin{bmatrix}
			a_{0(n-1)}&a_{0(n-2)}&\cdots&a_{01}&a_{00}\\
			a_{1(n-1)}&a_{1(n-2)}&\cdots&a_{11}&a_{10}\\
			\vdots&\vdots&\ddots&\vdots&\vdots\\
			a_{(m-1)(n-1)}&a_{(m-1)(n-2)}&\cdots&a_{(m-1)1}&a_{(m-1)0}
		\end{bmatrix}.$$
		Equivalently, the $(i,k)^{th}$ entry of $DG_{T_g^n}J_n$ is $(DG_{T_g^n}J_n)_{ik}
		=a_{i(n-1-k)}, 0\leq i\leq m-1$, $0\leq k\leq n-1.$ Since $f_i(x)=a_{i0}+a_{i1}x+\cdots+a_{i(n-1)}x^{n-1},$ therefore,
		by Definition \ref{length n reciprocal}, its length-$n$ reversal is $f_i^{*_{n}}(x)=x^{n-1}f_i(x^{-1}).$
		Substituting $x^{-1}$ into $f_i(x)$ gives $f_i(x^{-1})=a_{i0}+a_{i1}x^{-1}+\cdots
		+a_{i(n-1)}x^{-(n-1)}.$ Multiplying by $x^{n-1}$, we obtain
		$f_i^{*_{n}}(x)=x^{n-1}
		\left(
		a_{i0}
		+a_{i1}x^{-1}
		+\cdots
		+a_{i(n-1)}x^{-(n-1)}
		\right)\\
		=
		a_{i0}x^{n-1}
		+a_{i1}x^{n-2}
		+\cdots
		+a_{i(n-2)}x
		+a_{i(n-1)}.$
		Writing the polynomial in increasing powers of $x$, $f_i^{*_{n}}(x)=
		a_{i(n-1)}+a_{i(n-2)}x+\cdots+a_{i1}x^{n-2}+a_{i0}x^{n-1}.$ Hence the coefficient vector of $f_i^{*_{n}}(x)$ is $\operatorname{coeff}\bigl(f_i^{*_{n}}(x)\bigr)=
		\begin{bmatrix}
			a_{i(n-1)}&a_{i(n-2)}&\cdots&a_{i1}&a_{i0}
		\end{bmatrix}
		$ which is exactly the $i$-th row of $DG_{T_g^n}J_n$. Therefore, for every $i=0,1,\ldots,m-1$, $\operatorname{Row}_i(DG_{T_g^n}J_n)=
		\operatorname{coeff}\bigl(f_i^{*_{n}}(x)\bigr).$ Since $G_{T_{D,g}^n}$ is the matrix whose $i^{th}$ row is the coefficient vector of $f_i^{*_{n}}(x)$, it follows that $G_{T_{D,g}^n}=DG_{T_g^n}J_n.$
	\end{proof}
	
	\begin{lem}\label{linear1}\cite[Theorem 3.3]{Chaturvedi}
		Let $M$ be a $k\times n$ matrix over a finite commutative ring $R$ with unity such that $rank(M)=k$, $D$ be any $k\times k$ involutory matrix over $R$ and $M_D'$ be the associated $D-$ reflected matrix. Then the linear codes generated by $M$ and $M_D'$ have the same code parameters $[n,k,d].$ 
	\end{lem}
	\begin{lem}\label{linear2}\cite[Theorem 3.5]{Chaturvedi}
		Let $M$ be a $k\times n$ matrix over a finite commutative ring $R$ with unity such that $rank(M)=k$ and $D$ be any $k\times k$ involutory matrix over $R$. Then the linear code generated by the generator matrix $[M\mid M_D']$ is reversible.
	\end{lem}
	\begin{lem}\label{linear3}\cite[Theorem 3.7]{Chaturvedi}
		Let $M$ be a $k\times n$ matrix over a finite commutative ring $R$ with unity such that $rank(M)=k$ and $D$ be any $k\times k$ involutory matrix over $R$. Then the linear code generated by the generator matrix $[M\mid M_D']$ has length $2n$, size $|R^k|$ and distance greater than or equal to $2d$ where $n,k$ and $d$ are the code parameters of the linear code generated by $M$. 
	\end{lem}
	The following result gives the generalized $T_{D,g}^n-$set construction for reversible linear codes over a finite commmutative ring with unity for an arbitrary polynomial $g(x)$.
	\begin{thm}\label{linear theorem}
		Let $g(x)$ be a polynomial over a finite commutative ring with unity $R$ of $\deg(g(x))<n$. Let $D\in R^{m\times m}$ be an involutory matrix, where $m=n-\deg(g)$ and $T_{D,g}^n$ denote the $D$-generalized $T_n$-set associated with $g(x)$. Let $<T_g^n>$, $ <T_{D,g}^n>$ and $\langle T_g^n\mid T_{D,g}^n\rangle$ denote the linear codes generated by $G_{T_{g}^n}$, $G_{T_{D,g}^n}$ and $[G_{T_{g}^n}\mid G_{T_{D,g}^n}]$, respectively.
		Then the following statements hold:
		
		(1). The linear codes $<T_g^n>$ and $ <T_{D,g}^n>$ have the same $[n,k,d]$ parameters.
		
		(2). The code
		$C_D=\langle T_g^n\mid T_{D,g}^n\rangle$
		is reversible over $R$.
		
		(3). If $\langle T_g^n\rangle$ is an $[n,k,d]$ code, then $C_D$ has parameters $[2n,k,d']$, where
		$d'\geq 2d$.
		
	\end{thm}	
	\begin{proof}(1). By Lemma \ref{MainLemma}, $G_{T_{D,g}^n}$ is the $D-$reflected matrix of $G_{T_{g}^n}$. Therefore, if we take $M=G_{T_{g}^n}$ in Lemma \ref{linear1}, then $M_D'=G_{T_{D,g}^n}$. It follows that the linear codes generated by $G_{T_{g}^n}$ and $G_{T_{D,g}^n}$, that is, $<T_g^n>$ and $ <T_{D,g}^n>$ have the same $[n,k,d]$ parameters.\\
		
		(2). In Lemma \ref{linear2}, if we take $M=G_{T_{g}^n}$, then $M_D'=G_{T_{D,g}^n}$. Therefore,  for the linear code $\langle T_g^n\mid T_{D,g}^n\rangle$ the generator matrix is $[G_{T_{g}^n}\mid G_{T_{D,g}^n}]=[M\mid M_D']$. Hence, $\langle T_g^n\mid T_{D,g}^n\rangle$ is reversible.\\
		
		(3). In Lemma \ref{linear3}, if we take $M=G_{T_{g}^n}$, then $M_D'=G_{T_{D,g}^n}$. Therefore, for the linear code $\langle T_g^n\mid T_{D,g}^n\rangle$ the generator matrix is $[G_{T_{g}^n}\mid G_{T_{D,g}^n}]=[M\mid M_D']$. Hence $\langle T_g^n\mid T_{D,g}^n\rangle$ has code parameters $[2n,k,d']$, where $d'\geq 2d$.

	\end{proof}
	The following results from \cite{Oztas2} are consequences of Theorem \ref{linear theorem}, if we take the particular involutory matrix $D$ as the matrix $J_n$.
	\begin{coro}\label{linear corollary} (1).\cite[Theorem 1]{Oztas2} Let $g(x)$ be a polynomial over $F_q$ and $deg(g(x))<n.$ Then, the codes $<T_g^n>$ and $<T_{g^*}^n>$ have same $[n,k,d]$ parameters.
		
		(2).\cite[Theorem 2]{Oztas2} Let $g(x)$ be any polynomial over a commutative ring $R$ with $deg(g(x))<n$ and $g^*(x)$ be the reciprocal polynomial of $g(x).$ Then, $<T_g^n|T_{g^*}^n>$ is a reversible code over $R$, where the codes $<T_g^n>$ and $<T_{g^*}^n>$  have the same $[n,k,d]$ parameters. 
		
		(3).\cite[Theorem 3]{Oztas2} Let $g(x)$ be any polynomial over $F_{4^{2m}}$ with $deg(g(x))<n$ and $g^*(x)$ be the reciprocal polynomial of $g(x).$ Then, the reversible code $<T_g^n|T_{g^*}^n>$ has parameters $[2n,k,d_T]$, where $d_T\geq 2d$ and $<T_g^n>$ is a $[n,k,d]$ code.
	\end{coro}
	
	\begin{exl}\label{better linear parameter}
		
		Let $\mathbb{F}_{16}=\mathbb{F}_2(w)$, where $w$ is a primitive element of $\mathbb{F}_{16}$, satisfying $1+w+w^4=0$. Consider the polynomial $g(x)=(w^3+w^2+1)x^5+(w^3+w)x^4+wx^3+w^2x^2+x+1$. Here, $s=5$, $n=7$, and $m=n-s=2$. The reciprocal polynomial of $g(x)$ is $g^*(x)=x^5+x^4+w^2x^3+wx^2+(w^3+w)x+w^3+w^2+1$. The $T^n-$set construction of reversible linear code for the given polynomial in \cite[Appendix B]{Oztas} is an AMDS code.
		
		Take $D=\left(\begin{array}{cc}w^2&1\\w&w^2\end{array}\right)$. Using SageMath computations, $D^2=I_2$, so $D$ is involutory.
		The coefficient matrix associated with $g(x)$ is
		\[G_{T_g^7}=\left(\begin{array}{ccccccc}
			1&1&w^2&w&w^3+w&w^3+w^2+1&0\\
			0&1&1&w^2&w&w^3+w&w^3+w^2+1
		\end{array}.\right)\]
		The generator matrix corresponding to $T_{D,g}^{7}$ is,
		\begin{tiny}
			\[G_{T_{D,g}^{7}}=\left(\begin{array}{ccccccc}
				w^3+w^2+1&w^3+w&w^3+w^2&w^3+w^2+1&w^3+w^2+w&w^2+1&w\\
				w^2+w+1&w^3+w^2+w+1&w^3+w^2+w+1&w^2+w+1&w^2&w&w^2
			\end{array}\right).\]
		\end{tiny}
		By SageMath computations, the first block code and the second block code both have parameters $[7,2,6]$, while the concatenated code $\langle T_g^7\mid T_{D,g}^7\rangle$ has parameters $[14,2,13]$. Hence, its Singleton defect is $\delta=14-2+1-13=0$, and the concatenated code is a reversible linear MDS code. The number of codewords is $16^2=256$.
		
	\end{exl}
	
	\begin{rmk}\label{paper 3 mds amds number}
		Consider $g(x)=w+x+w^2x^2+w^3x^3$, with $n=5$ and $s=3$ over $F_{16}$. Then the coefficient matrix 		$G_{T_g^5}=\begin{pmatrix}
			w&1&w^2&w^3&0\\
			0&w&1&w^2&w^3
		\end{pmatrix}$. By SageMath computations \cite{Sage}, the $T^n-$set construction gives an AMDS code. Since there are $256$ involutory matrices of order $2\times 2$ over the finite field $F_{16}$, we have $256$ choices for the $T_{D,g}^n$ construction. Out of $256$ reversible linear codes obtained, $181$ are AMDS and $75$ are MDS codes. Therefore, $T_{D,g}^n$ construction not only broadens the scope of construction in terms of number of generator matrices for obtaining reversible linear codes but it also yields MDS codes.		
		
	\end{rmk}
	Here, we use the following table of DNA correspondence over $\mathbb{F}_{16}$ developed by Oztas and Siap \cite{Siap}, also mentioned in \cite[Table 2]{Oztas}.

	\begin{table}[h]\label{table 1}
		\centering

		\caption{DNA correspondence over $\mathbb{F}_{16},$ \cite[Table 2]{Oztas}} \label{DNA table}
		\begin{tabular}{|c|c|c|c|} \hline
			
			Index & DNA Pair & $\mathbb{F}_{16}$ (Multiplicative ) & Additive Form \\ \hline
			
			1  & AA & $0$       & $0$ \\
			2  & TT & $w^0$     & $1$ \\
			3  & AT & $w^1$     & $w$ \\
			4  & GC & $w^2$     & $w^2$ \\
			5  & AG & $w^3$     & $w^3$ \\
			6  & TA & $w^4$     & $1+w$ \\
			7  & CC & $w^5$     & $w+w^2$ \\
			8  & AC & $w^6$     & $w^2+w^3$ \\
			9  & GT & $w^7$     & $1+w+w^3$ \\
			10 & CG & $w^8$     & $1+w^2$ \\
			11 & CA & $w^9$     & $w+w^3$ \\
			12 & GG & $w^{10}$  & $1+w+w^2$ \\
			13 & CT & $w^{11}$  & $w+w^2+w^3$ \\
			14 & GA & $w^{12}$  & $1+w+w^2+w^3$ \\
			15 & TG & $w^{13}$  & $1+w^2+w^3$ \\
			16 & TC & $w^{14}$  & $1+w^3$ \\ \hline
			
		\end{tabular}
	\end{table}

	 We recall the key observations from \cite[Section IV]{Oztas}. Let $\Phi$ be the one-to-one mapping that defines the DNA representation for elements within $\mathbb{F}_{16}$. For example, we are given that $\Phi(w^{11}) = \text{CT}$. Adding any field element to $1$ corresponds to finding the complement of its DNA equivalent, meaning $\Phi(w^{11} + 1) = \Phi(u^{12}) = \text{GA}$. Moreover, computing the fourth power of a field element reverses the order of its corresponding DNA pairs. For example $\Phi((w^{11})^4) = \Phi(w^{44}) = \text{TC}$. Finally, this algebraic mapping $\Phi$ extends naturally to codewords and codes. We recall the following results from \cite{Chaturvedi}.
	
	\begin{lem}\cite[Corollary 5.5]{Chaturvedi}\label{DNA 1} Let $M$ be a $k\times n$ matrix over a finite field $\mathbb{F}_{4^{2t}}$ such that $rank(M)=k$ and $D$ be any $k\times k$ involutory matrix over $\mathbb{F}_{4^{2t}}$. Then the linear codes generated by $M$ and $(M_D')^{o4^t}$ have the same code parameters $[n,k,d].$
	\end{lem}
	
	\begin{thm}\cite[Theorem 5.9]{Chaturvedi}\label{DNA 2} Let $M$ be a $k\times n$ matrix over a field $F_{4^{2t}}$ such that $rank(M)=k$ and $D$ be any $k\times k$ involutory matrix over $F_{4^{2t}}$ such that $D^{o4^t}=D.$ Let $C$ be the linear code generated by $[M\mid (M'_D)^{o4^t}]$. Then the code $\Phi(C)$ is reversible. Moreover, the code parameters of $C$ are $2n,\:k,\:d'$ where $d'\geq 2d$ and $n,k,d$ are the code parameters of the linear code generated by $M.$
	\end{thm}
	
	We give the following result for construction of reversible DNA codes using $T_{D,g}^n-$set.
	\begin{thm}\label{DNA theorem}
		Let $g(x)$ be a polynomial over a finite finite field $F_{4^{2t}}$ where $t$ is a natural number, with $\deg(g(x))<n$. Let $D\in R^{m\times m}$ be an involutory matrix, where $m=n-\deg(g)$ such that $D^{o4^t}=D.$ Let $T_{D,g}^n$ denote the $D$-generalized $T_n$-set associated with $g(x)$.
		Then the following statements hold:
		
		(1). The linear codes $<T_g^n>$ and $ <(T_{D,g}^n)^{o4^t}>$ have the same $[n,k,d]$ parameters.
		
		(2). The DNA code
		$\Phi(\langle T_g^n\mid (T_{D,g}^n)^{o4^t}\rangle)$ is a reversible DNA code.
		
		(3). If $\langle T_g^n\rangle$ is an $[n,k,d]$ code, then $\langle T_g^n\mid (T_{D,g}^n)^{o4^t}\rangle$ has parameters $[2n,k,d']$, where $d'\geq 2d$.
	\end{thm}
	\begin{proof}
		(1). By Lemma \ref{MainLemma}, $G_{T_{D,g}^n}$ is the $D-$reflected matrix of $G_{T_{g}^n}$. Therefore, in Lemma \ref{DNA 1}, if we take $M=G_{T_{g}^n}$, then $M_D'=G_{T_{D,g}^n}$. It follows that the linear codes generated by $G_{T_{g}^n}$ and $G_{T_{D,g}^n}^{o4^t}$, that is, $<T_g^n>$ and $ <(T_{D,g}^n)^{o4^t}>$ have the same $[n,k,d]$ parameters.\\
		
		(2). In Theorem \ref{DNA 2}, if we take $M=G_{T_{g}^n}$, then by Lemma \ref{MainLemma}, $M_D'=G_{T_{D,g}^n}$. Therefore, the generator matrix for the linear code $\langle T_g^n\mid (T_{D,g}^n)^{o4^t}\rangle$, that is, $[G_{T_{g}^n}\mid G_{T_{D,g}^n}^{o4^t}]$ is equal to $[M\mid M_D'^{o4^t}]$. Hence, the DNA code	$\Phi(\langle T_g^n\mid (T_{D,g}^n)^{o4^t}\rangle)$ is a reversible DNA code.\\
		
		(3). In Theorem \ref{DNA 2}, if we take $M=G_{T_{g}^n}$, then by Lemma \ref{MainLemma}, $M_D'=G_{T_{D,g}^n}$. Moreover, the linear code generated by $M$ has code parameters $[n,k,d].$ The generator matrix for the linear code $\langle T_g^n\mid (T_{D,g}^n)^{o4^t}\rangle$, that is, $[G_{T_{g}^n}\mid G_{T_{D,g}^n}^{o4^t}]$ is equal to $[M\mid M_D'^{o4^t}]$. Hence, the linear code	$\langle T_g^n\mid (T_{D,g}^n)^{o4^t}\rangle$ has parameters $[2n,k,d']$, where $d'\geq 2d$.
	\end{proof}
	
	The following results from \cite{Oztas2} are direct consequences of Theorem \ref{DNA theorem} by taking the involutory matrix $D$ as $J_n.$
	
	\begin{coro}\label{DNA corollary} (1). \cite[Corollary 6]{Oztas2} Let $g(x)$ be a polynomial over $F_{4^{2m}}$ and $deg(g(x))<n.$ Then, codes $<T_g^n>$ and $<T_{({g^*})^{o4^m}}^n>$ have the same $[n,k,d]$ parameters.
		
		(2). \cite[Theorem 7]{Oztas2} Let $g(x)$ be any polynomial over $F_{4^{2m}}$ with $deg(g(x))<n$ and $g^*(x)$ be the reciprocal polynomial of $g(x).$ Then, the code $C'=<T_g^n|T_{({g^*})^{o4^m}}^n>$ has $[2n,k,d_T]$  parameters, and $\Phi(C')$ is a reversible DNA code, where $d_T\geq 2d$, $<T_g^n>$ is a $[n,k,d]$ code.
	\end{coro}
	\begin{exl}\label{DNA Example}
		Let $F=\mathbb{F}_{16}=\mathbb{F}_2(w)$, where $w^4+w+1=0$. Consider the polynomial
		$g(x)=w+x+w^2x^2+w^3x^3$, with $n=5$ and $s=3$.
		
		The coefficient matrix is
		$G_{T_g^5}=\begin{pmatrix}
			w&1&w^2&w^3&0\\
			0&w&1&w^2&w^3
		\end{pmatrix}$.			
		Consequently,
		$G_{T_g^5}J_5$$=\begin{pmatrix}
			0&w^3&w^2&1&w\\
			w^3&w^2&1&w&0
		\end{pmatrix}$.	Let $D=\begin{pmatrix}
			w^2+w+1&w^2+w\\
			w^2+w&w^2+w+1
		\end{pmatrix}
		=\begin{pmatrix}
			w^{10}&w^5\\
			w^5&w^{10}
		\end{pmatrix}$.			
		By Sage Math computations, $D^2=I_d$ and $D^{o4}=D$, where $D^{o4}$ denotes the Hadamard fourth power of $D$.			
		Multiplying $D$ by $G_{T_g^5}J_5$ gives
		$DG_{T_g^5}J_5$$=\begin{pmatrix}
			w^8&w^5&w^{14}&w^7&w^{11}\\
			w^{13}&w^9&w^6&w^3&w^6
		\end{pmatrix}$.			
		Taking the Hadamard fourth power, we obtain
		$(DG_{T_g^n}J_5)^{o4}$$=\begin{pmatrix}
			w^2&w^5&w^{11}&w^{13}&w^{14}\\
			w^7&w^6&w^9&w^{12}&w^9
		\end{pmatrix}$,
		where exponents are reduced modulo $15$, since $w^{15}=1$.
		
		Therefore, the concatenated generator matrix
		$[G_{T_g^5}|(G_{T_{D,g}^5})^{o4}]$$=\left[G_{T_g^5}\mid (DG_{T_g^5}J_5)^{o4}\right]$
		is
		\[[G_{T_g^5}|(G_{T_{D,g}^5})^{o4}]=\begin{pmatrix}
			w&1&w^2&w^3&0&w^2&w^5&w^{11}&w^{13}&w^{14}\\
			0&w&1&w^2&w^3&w^7&w^6&w^9&w^{12}&w^9
		\end{pmatrix}.\] The DNA code corresponding to the linear code generated by this matrix is a reversible DNA code. Moreover, the $[n,k,d]$ parameters are $[10,2,9]$ making it an MDS code. However, if we take $D=J_2$ which corresponds to the $T_g^5$ construction the $[n,k,d]$ parameters are $[10,2,8]$, that is, the obtained code is AMDS in this case.
		
		Table \ref{DNA table 2} contains explicitly the generated DNA code. 
		\footnotesize
		\begin{longtable}{|c|l|l|}
			\caption{DNA correspondence table for the Example \ref{DNA Example}}\label{DNA table 2}\\
			\hline
			No. & Field Codeword & DNA \\
			\hline
			\endfirsthead
			\hline
			No. & Field Codeword & DNA \\
			\hline
			\endhead
			1 & $0 0 0 0 0 0 0 0 0 0$ & \texttt{AAAAAAAAAAAAAAAAAAAA} \\
			\hline
			2 & $w^{1} 1 w^{2} w^{3} 0 w^{2} w^{5} w^{11} w^{13} w^{14}$ & \texttt{ATTTGCAGAAGCCCCTTGTC} \\
			\hline
			3 & $w^{2} w^{1} w^{3} w^{4} 0 w^{3} w^{6} w^{12} w^{14} 1$ & \texttt{GCATAGTAAAAGACGATCTT} \\
			\hline
			4 & $w^{5} w^{4} w^{6} w^{7} 0 w^{6} w^{9} 1 w^{2} w^{3}$ & \texttt{CCTAACGTAAACCATTGCAG} \\
			\hline
			5 & $w^{3} w^{2} w^{4} w^{5} 0 w^{4} w^{7} w^{13} 1 w^{1}$ & \texttt{AGGCTACCAATAGTTGTTAT} \\
			\hline
			6 & $w^{9} w^{8} w^{10} w^{11} 0 w^{10} w^{13} w^{4} w^{6} w^{7}$ & \texttt{CACGGGCTAAGGTGTAACGT} \\
			\hline
			7 & $w^{6} w^{5} w^{7} w^{8} 0 w^{7} w^{10} w^{1} w^{3} w^{4}$ & \texttt{ACCCGTCGAAGTGGATAGTA} \\
			\hline
			8 & $w^{11} w^{10} w^{12} w^{13} 0 w^{12} 1 w^{6} w^{8} w^{9}$ & \texttt{CTGGGATGAAGATTACCGCA} \\
			\hline
			9 & $w^{4} w^{3} w^{5} w^{6} 0 w^{5} w^{8} w^{14} w^{1} w^{2}$ & \texttt{TAAGCCACAACCCGTCATGC} \\
			\hline
			10 & $1 w^{14} w^{1} w^{2} 0 w^{1} w^{4} w^{10} w^{12} w^{13}$ & \texttt{TTTCATGCAAATTAGGGATG} \\
			\hline
			11 & $w^{10} w^{9} w^{11} w^{12} 0 w^{11} w^{14} w^{5} w^{7} w^{8}$ & \texttt{GGCACTGAAACTTCCCGTCG} \\
			\hline
			12 & $w^{8} w^{7} w^{9} w^{10} 0 w^{9} w^{12} w^{3} w^{5} w^{6}$ & \texttt{CGGTCAGGAACAGAAGCCAC} \\
			\hline
			13 & $w^{7} w^{6} w^{8} w^{9} 0 w^{8} w^{11} w^{2} w^{4} w^{5}$ & \texttt{GTACCGCAAACGCTGCTACC} \\
			\hline
			14 & $w^{14} w^{13} 1 w^{1} 0 1 w^{3} w^{9} w^{11} w^{12}$ & \texttt{TCTGTTATAATTAGCACTGA} \\
			\hline
			15 & $w^{12} w^{11} w^{13} w^{14} 0 w^{13} w^{1} w^{7} w^{9} w^{10}$ & \texttt{GACTTGTCAATGATGTCAGG} \\
			\hline
			16 & $w^{13} w^{12} w^{14} 1 0 w^{14} w^{2} w^{8} w^{10} w^{11}$ & \texttt{TGGATCTTAATCGCCGGGCT} \\
			\hline
			17 & $0 w^{1} 1 w^{2} w^{3} w^{7} w^{6} w^{9} w^{12} w^{9}$ & \texttt{AAATTTGCAGGTACCAGACA} \\
			\hline
			18 & $w^{1} w^{4} w^{8} w^{6} w^{3} w^{12} w^{9} w^{2} w^{1} w^{4}$ & \texttt{ATTACGACAGGACAGCATTA} \\
			\hline
			19 & $w^{2} 0 w^{14} w^{10} w^{3} w^{4} 0 w^{8} w^{5} w^{7}$ & \texttt{GCAATCGGAGTAAACGCCGT} \\
			\hline
			20 & $w^{5} 1 w^{13} w^{12} w^{3} w^{10} w^{5} w^{7} w^{7} w^{1}$ & \texttt{CCTTTGGAAGGGCCGTGTAT} \\
			\hline
			21 & $w^{3} w^{5} w^{1} w^{1} w^{3} w^{3} w^{10} w^{10} w^{11} w^{3}$ & \texttt{AGCCATATAGAGGGGGCTAG} \\
			\hline
			22 & $w^{9} w^{10} w^{5} w^{9} w^{3} w^{6} 1 w^{14} w^{4} 1$ & \texttt{CAGGCCCAAGACTTTCTATT} \\
			\hline
			23 & $w^{6} w^{2} w^{9} 1 w^{3} 0 w^{7} w^{3} w^{10} w^{14}$ & \texttt{ACGCCATTAGAAGTAGGGTC} \\
			\hline
			24 & $w^{11} w^{8} w^{11} w^{14} w^{3} w^{2} w^{13} w^{5} w^{9} 0$ & \texttt{CTCGCTTCAGGCTGCCCAAA} \\
			\hline
			25 & $w^{4} w^{9} w^{10} w^{3} w^{3} w^{13} w^{14} w^{4} w^{13} w^{11}$ & \texttt{TACAGGAGAGTGTCTATGCT} \\
			\hline
			26 & $1 w^{7} w^{4} 0 w^{3} w^{14} w^{12} w^{13} 0 w^{10}$ & \texttt{TTGTTAAAAGTCGATGAAGG} \\
			\hline
			27 & $w^{10} w^{3} w^{12} w^{7} w^{3} w^{8} w^{8} w^{6} w^{2} w^{12}$ & \texttt{GGAGGAGTAGCGCGACGCGA} \\
			\hline
			28 & $w^{8} w^{14} w^{7} w^{4} w^{3} 1 w^{4} w^{1} w^{14} w^{5}$ & \texttt{CGTCGTTAAGTTTAATTCCC} \\
			\hline
			29 & $w^{7} w^{11} w^{2} w^{11} w^{3} w^{11} w^{1} w^{11} w^{6} w^{6}$ & \texttt{GTCTGCCTAGCTATCTACAC} \\
			\hline
			30 & $w^{14} w^{12} 0 w^{5} w^{3} w^{9} w^{2} 0 1 w^{8}$ & \texttt{TCGAAACCAGCAGCAATTCG} \\
			\hline
			31 & $w^{12} w^{6} w^{6} w^{13} w^{3} w^{5} w^{11} 1 w^{8} w^{13}$ & \texttt{GAACACTGAGCCCTTTCGTG} \\
			\hline
			32 & $w^{13} w^{13} w^{3} w^{8} w^{3} w^{1} w^{3} w^{12} w^{3} w^{2}$ & \texttt{TGTGAGCGAGATAGGAAGGC} \\
			\hline
			33 & $0 w^{2} w^{1} w^{3} w^{4} w^{8} w^{7} w^{10} w^{13} w^{10}$ & \texttt{AAGCATAGTACGGTGGTGGG} \\
			\hline
			34 & $w^{1} w^{8} w^{5} 0 w^{4} 1 w^{13} w^{14} 0 w^{11}$ & \texttt{ATCGCCAATATTTGTCAACT} \\
			\hline
			35 & $w^{2} w^{5} w^{9} w^{7} w^{4} w^{13} w^{10} w^{3} w^{2} w^{5}$ & \texttt{GCCCCAGTTATGGGAGGCCC} \\
			\hline
			36 & $w^{5} w^{10} w^{11} w^{4} w^{4} w^{14} 1 w^{5} w^{14} w^{12}$ & \texttt{CCGGCTTATATCTTCCTCGA} \\
			\hline
			37 & $w^{3} 0 1 w^{11} w^{4} w^{5} 0 w^{9} w^{6} w^{8}$ & \texttt{AGAATTCTTACCAACAACCG} \\
			\hline
			38 & $w^{9} 1 w^{8} w^{5} w^{4} w^{1} w^{5} w^{2} 1 w^{6}$ & \texttt{CATTCGCCTAATCCGCTTAC} \\
			\hline
			39 & $w^{6} w^{1} w^{14} w^{13} w^{4} w^{11} w^{6} w^{8} w^{8} w^{2}$ & \texttt{ACATTCTGTACTACCGCGGC} \\
			\hline
			40 & $w^{11} w^{4} w^{13} w^{8} w^{4} w^{9} w^{9} w^{7} w^{3} w^{13}$ & \texttt{CTTATGCGTACACAGTAGTG} \\
			\hline
			41 & $w^{4} w^{6} w^{2} w^{2} w^{4} w^{4} w^{11} w^{11} w^{12} w^{4}$ & \texttt{TAACGCGCTATACTCTGATA} \\
			\hline
			42 & $1 w^{13} 0 w^{6} w^{4} w^{10} w^{3} 0 w^{1} w^{9}$ & \texttt{TTTGAAACTAGGAGAAATCA} \\
			\hline
			43 & $w^{10} w^{11} w^{6} w^{10} w^{4} w^{7} w^{1} 1 w^{5} w^{1}$ & \texttt{GGCTACGGTAGTATTTCCAT} \\
			\hline
			44 & $w^{8} w^{12} w^{3} w^{12} w^{4} w^{12} w^{2} w^{12} w^{7} w^{7}$ & \texttt{CGGAAGGATAGAGCGAGTGT} \\
			\hline
			45 & $w^{7} w^{3} w^{10} w^{1} w^{4} 0 w^{8} w^{4} w^{11} 1$ & \texttt{GTAGGGATTAAACGTACTTT} \\
			\hline
			46 & $w^{14} w^{14} w^{4} w^{9} w^{4} w^{2} w^{4} w^{13} w^{4} w^{3}$ & \texttt{TCTCTACATAGCTATGTAAG} \\
			\hline
			47 & $w^{12} w^{9} w^{12} 1 w^{4} w^{3} w^{14} w^{6} w^{10} 0$ & \texttt{GACAGATTTAAGTCACGGAA} \\
			\hline
			48 & $w^{13} w^{7} w^{7} w^{14} w^{4} w^{6} w^{12} w^{1} w^{9} w^{14}$ & \texttt{TGGTGTTCTAACGAATCATC} \\
			\hline
			49 & $0 w^{5} w^{4} w^{6} w^{7} w^{11} w^{10} w^{13} w^{1} w^{13}$ & \texttt{AACCTAACGTCTGGTGATTG} \\
			\hline
			50 & $w^{1} w^{10} w^{10} w^{2} w^{7} w^{9} 1 w^{4} w^{12} w^{2}$ & \texttt{ATGGGGGCGTCATTTAGAGC} \\
			\hline
			51 & $w^{2} w^{2} w^{7} w^{12} w^{7} w^{5} w^{7} w^{1} w^{7} w^{6}$ & \texttt{GCGCGTGAGTCCGTATGTAC} \\
			\hline
			52 & $w^{5} w^{8} w^{12} w^{10} w^{7} w^{1} w^{13} w^{6} w^{5} w^{8}$ & \texttt{CCCGGAGGGTATTGACCCCG} \\
			\hline
			53 & $w^{3} w^{1} 0 w^{9} w^{7} w^{13} w^{6} 0 w^{4} w^{12}$ & \texttt{AGATAACAGTTGACAATAGA} \\
			\hline
			54 & $w^{9} w^{4} w^{2} w^{1} w^{7} w^{14} w^{9} w^{11} w^{11} w^{5}$ & \texttt{CATAGCATGTTCCACTCTCC} \\
			\hline
			55 & $w^{6} 0 w^{3} w^{14} w^{7} w^{8} 0 w^{12} w^{9} w^{11}$ & \texttt{ACAAAGTCGTCGAAGACACT} \\
			\hline
			56 & $w^{11} 1 w^{6} 1 w^{7} 1 w^{5} 1 w^{10} w^{10}$ & \texttt{CTTTACTTGTTTCCTTGGGG} \\
			\hline
			57 & $w^{4} w^{11} w^{8} 0 w^{7} w^{3} w^{1} w^{2} 0 w^{14}$ & \texttt{TACTCGAAGTAGATGCAATC} \\
			\hline
			58 & $1 w^{12} 1 w^{3} w^{7} w^{6} w^{2} w^{9} w^{13} 0$ & \texttt{TTGATTAGGTACGCCATGAA} \\
			\hline
			59 & $w^{10} w^{6} w^{13} w^{4} w^{7} 0 w^{11} w^{7} w^{14} w^{3}$ & \texttt{GGACTGTAGTAACTGTTCAG} \\
			\hline
			60 & $w^{8} w^{13} w^{14} w^{7} w^{7} w^{2} w^{3} w^{8} w^{2} 1$ & \texttt{CGTGTCGTGTGCAGCGGCTT} \\
			\hline
			61 & $w^{7} w^{9} w^{5} w^{5} w^{7} w^{7} w^{14} w^{14} 1 w^{7}$ & \texttt{GTCACCCCGTGTTCTCTTGT} \\
			\hline
			62 & $w^{14} w^{7} w^{1} w^{11} w^{7} w^{12} w^{12} w^{10} w^{6} w^{1}$ & \texttt{TCGTATCTGTGAGAGGACAT} \\
			\hline
			63 & $w^{12} w^{3} w^{11} w^{8} w^{7} w^{4} w^{8} w^{5} w^{3} w^{9}$ & \texttt{GAAGCTCGGTTACGCCAGCA} \\
			\hline
			64 & $w^{13} w^{14} w^{9} w^{13} w^{7} w^{10} w^{4} w^{3} w^{8} w^{4}$ & \texttt{TGTCCATGGTGGTAAGCGTA} \\
			\hline
			65 & $0 w^{3} w^{2} w^{4} w^{5} w^{9} w^{8} w^{11} w^{14} w^{11}$ & \texttt{AAAGGCTACCCACGCTTCCT} \\
			\hline
			66 & $w^{1} w^{14} 0 w^{7} w^{5} w^{11} w^{4} 0 w^{2} w^{10}$ & \texttt{ATTCAAGTCCCTTAAAGCGG} \\
			\hline
			67 & $w^{2} w^{9} w^{6} 0 w^{5} w^{1} w^{14} 1 0 w^{12}$ & \texttt{GCCAACAACCATTCTTAAGA} \\
			\hline
			68 & $w^{5} w^{7} w^{3} w^{3} w^{5} w^{5} w^{12} w^{12} w^{13} w^{5}$ & \texttt{CCGTAGAGCCCCGAGATGCC} \\
			\hline
			69 & $w^{3} w^{6} w^{10} w^{8} w^{5} w^{14} w^{11} w^{4} w^{3} w^{6}$ & \texttt{AGACGGCGCCTCCTTAAGAC} \\
			\hline
			70 & $w^{9} w^{13} w^{4} w^{13} w^{5} w^{13} w^{3} w^{13} w^{8} w^{8}$ & \texttt{CATGTATGCCTGAGTGCGCG} \\
			\hline
			71 & $w^{6} w^{11} w^{12} w^{5} w^{5} 1 w^{1} w^{6} 1 w^{13}$ & \texttt{ACCTGACCCCTTATACTTTG} \\
			\hline
			72 & $w^{11} w^{12} w^{7} w^{11} w^{5} w^{8} w^{2} w^{1} w^{6} w^{2}$ & \texttt{CTGAGTCTCCCGGCATACGC} \\
			\hline
			73 & $w^{4} 0 w^{1} w^{12} w^{5} w^{6} 0 w^{10} w^{7} w^{9}$ & \texttt{TAAAATGACCACAAGGGTCA} \\
			\hline
			74 & $1 1 w^{5} w^{10} w^{5} w^{3} w^{5} w^{14} w^{5} w^{4}$ & \texttt{TTTTCCGGCCAGCCTCCCTA} \\
			\hline
			75 & $w^{10} w^{1} w^{9} w^{6} w^{5} w^{2} w^{6} w^{3} w^{1} w^{7}$ & \texttt{GGATCAACCCGCACAGATGT} \\
			\hline
			76 & $w^{8} w^{4} w^{11} w^{2} w^{5} 0 w^{9} w^{5} w^{12} w^{1}$ & \texttt{CGTACTGCCCAACACCGAAT} \\
			\hline
			77 & $w^{7} w^{2} 1 w^{14} w^{5} w^{12} w^{7} w^{9} w^{9} w^{3}$ & \texttt{GTGCTTTCCCGAGTCACAAG} \\
			\hline
			78 & $w^{14} w^{8} w^{8} 1 w^{5} w^{7} w^{13} w^{2} w^{10} 1$ & \texttt{TCCGCGTTCCGTTGGCGGTT} \\
			\hline
			79 & $w^{12} w^{5} w^{14} w^{9} w^{5} w^{10} w^{10} w^{8} w^{4} w^{14}$ & \texttt{GACCTCCACCGGGGCGTATC} \\
			\hline
			80 & $w^{13} w^{10} w^{13} w^{1} w^{5} w^{4} 1 w^{7} w^{11} 0$ & \texttt{TGGGTGATCCTATTGTCTAA} \\
			\hline
			81 & $0 w^{9} w^{8} w^{10} w^{11} 1 w^{14} w^{2} w^{5} w^{2}$ & \texttt{AACACGGGCTTTTCGCCCGC} \\
			\hline
			82 & $w^{1} w^{7} 1 w^{12} w^{11} w^{8} w^{12} w^{9} w^{7} w^{13}$ & \texttt{ATGTTTGACTCGGACAGTTG} \\
			\hline
			83 & $w^{2} w^{3} w^{13} w^{2} w^{11} w^{14} w^{8} w^{7} w^{12} w^{8}$ & \texttt{GCAGTGGCCTTCCGGTGACG} \\
			\hline
			84 & $w^{5} w^{14} w^{14} w^{6} w^{11} w^{13} w^{4} w^{8} w^{1} w^{6}$ & \texttt{CCTCTCACCTTGTACGATAC} \\
			\hline
			85 & $w^{3} w^{11} w^{5} 1 w^{11} w^{1} w^{1} w^{14} w^{10} w^{5}$ & \texttt{AGCTCCTTCTATATTCGGCC} \\
			\hline
			86 & $w^{9} w^{12} w^{1} w^{14} w^{11} w^{5} w^{2} w^{10} w^{9} w^{12}$ & \texttt{CAGAATTCCTCCGCGGCAGA} \\
			\hline
			87 & $w^{6} w^{6} w^{11} w^{1} w^{11} w^{9} w^{11} w^{5} w^{11} w^{10}$ & \texttt{ACACCTATCTCACTCCCTGG} \\
			\hline
			88 & $w^{11} w^{13} w^{9} w^{9} w^{11} w^{11} w^{3} w^{3} w^{4} w^{11}$ & \texttt{CTTGCACACTCTAGAGTACT} \\
			\hline
			89 & $w^{4} w^{1} w^{4} w^{7} w^{11} w^{10} w^{6} w^{13} w^{2} 0$ & \texttt{TAATTAGTCTGGACTGGCAA} \\
			\hline
			90 & $1 w^{4} w^{10} w^{4} w^{11} w^{4} w^{9} w^{4} w^{14} w^{14}$ & \texttt{TTTAGGTACTTACATATCTC} \\
			\hline
			91 & $w^{10} 0 w^{7} w^{3} w^{11} w^{12} 0 w^{1} w^{13} 1$ & \texttt{GGAAGTAGCTGAAAATTGTT} \\
			\hline
			92 & $w^{8} 1 w^{12} 0 w^{11} w^{7} w^{5} w^{6} 0 w^{3}$ & \texttt{CGTTGAAACTGTCCACAAAG} \\
			\hline
			93 & $w^{7} w^{5} 0 w^{13} w^{11} w^{2} w^{10} 0 w^{8} w^{1}$ & \texttt{GTCCAATGCTGCGGAACGAT} \\
			\hline
			94 & $w^{14} w^{10} w^{2} w^{8} w^{11} 0 1 w^{11} w^{3} w^{7}$ & \texttt{TCGGGCCGCTAATTCTAGGT} \\
			\hline
			95 & $w^{12} w^{2} w^{3} w^{11} w^{11} w^{6} w^{7} w^{12} w^{6} w^{4}$ & \texttt{GAGCAGCTCTACGTGAACTA} \\
			\hline
			96 & $w^{13} w^{8} w^{6} w^{5} w^{11} w^{3} w^{13} 1 1 w^{9}$ & \texttt{TGCGACCCCTAGTGTTTTCA} \\
			\hline
			97 & $0 w^{6} w^{5} w^{7} w^{8} w^{12} w^{11} w^{14} w^{2} w^{14}$ & \texttt{AAACCCGTCGGACTTCGCTC} \\
			\hline
			98 & $w^{1} w^{13} w^{1} w^{4} w^{8} w^{7} w^{3} w^{10} w^{14} 0$ & \texttt{ATTGATTACGGTAGGGTCAA} \\
			\hline
			99 & $w^{2} w^{11} w^{11} w^{3} w^{8} w^{10} w^{1} w^{5} w^{13} w^{3}$ & \texttt{GCCTCTAGCGGGATCCTGAG} \\
			\hline
			100 & $w^{5} w^{12} w^{9} 0 w^{8} w^{4} w^{2} w^{3} 0 1$ & \texttt{CCGACAAACGTAGCAGAATT} \\
			\hline
			101 & $w^{3} w^{3} w^{8} w^{13} w^{8} w^{6} w^{8} w^{2} w^{8} w^{7}$ & \texttt{AGAGCGTGCGACCGGCCGGT} \\
			\hline
			102 & $w^{9} w^{14} 1 w^{8} w^{8} w^{3} w^{4} w^{9} w^{3} w^{1}$ & \texttt{CATCTTCGCGAGTACAAGAT} \\
			\hline
			103 & $w^{6} w^{9} w^{13} w^{11} w^{8} w^{2} w^{14} w^{7} w^{6} w^{9}$ & \texttt{ACCATGCTCGGCTCGTACCA} \\
			\hline
			104 & $w^{11} w^{7} w^{14} w^{5} w^{8} 0 w^{12} w^{8} 1 w^{4}$ & \texttt{CTGTTCCCCGAAGACGTTTA} \\
			\hline
			105 & $w^{4} w^{2} 0 w^{10} w^{8} w^{14} w^{7} 0 w^{5} w^{13}$ & \texttt{TAGCAAGGCGTCGTAACCTG} \\
			\hline
			106 & $1 w^{8} w^{2} w^{12} w^{8} w^{13} w^{13} w^{11} w^{7} w^{2}$ & \texttt{TTCGGCGACGTGTGCTGTGC} \\
			\hline
			107 & $w^{10} w^{5} w^{3} w^{2} w^{8} 1 w^{10} w^{12} w^{12} w^{6}$ & \texttt{GGCCAGGCCGTTGGGAGAAC} \\
			\hline
			108 & $w^{8} w^{10} w^{6} w^{6} w^{8} w^{8} 1 1 w^{1} w^{8}$ & \texttt{CGGGACACCGCGTTTTATCG} \\
			\hline
			109 & $w^{7} 0 w^{4} 1 w^{8} w^{9} 0 w^{13} w^{10} w^{12}$ & \texttt{GTAATATTCGCAAATGGGGA} \\
			\hline
			110 & $w^{14} 1 w^{10} w^{14} w^{8} w^{11} w^{5} w^{4} w^{9} w^{5}$ & \texttt{TCTTGGTCCGCTCCTACACC} \\
			\hline
			111 & $w^{12} w^{1} w^{7} w^{1} w^{8} w^{1} w^{6} w^{1} w^{11} w^{11}$ & \texttt{GAATGTATCGATACATCTCT} \\
			\hline
			112 & $w^{13} w^{4} w^{12} w^{9} w^{8} w^{5} w^{9} w^{6} w^{4} w^{10}$ & \texttt{TGTAGACACGCCCAACTAGG} \\
			\hline
			113 & $0 w^{11} w^{10} w^{12} w^{13} w^{2} w^{1} w^{4} w^{7} w^{4}$ & \texttt{AACTGGGATGGCATTAGTTA} \\
			\hline
			114 & $w^{1} w^{12} w^{4} w^{10} w^{13} 0 w^{2} w^{13} w^{5} w^{9}$ & \texttt{ATGATAGGTGAAGCTGCCCA} \\
			\hline
			115 & $w^{2} w^{6} w^{12} w^{6} w^{13} w^{6} w^{11} w^{6} w^{1} w^{1}$ & \texttt{GCACGAACTGACCTACATAT} \\
			\hline
			116 & $w^{5} w^{13} w^{7} w^{2} w^{13} w^{3} w^{3} w^{1} w^{12} w^{7}$ & \texttt{CCTGGTGCTGAGAGATGAGT} \\
			\hline
			117 & $w^{3} w^{9} w^{2} w^{14} w^{13} w^{10} w^{14} w^{11} w^{9} 1$ & \texttt{AGCAGCTCTGGGTCCTCATT} \\
			\hline
			118 & $w^{9} w^{7} 0 1 w^{13} w^{4} w^{12} 0 w^{10} w^{3}$ & \texttt{CAGTAATTTGTAGAAAGGAG} \\
			\hline
			119 & $w^{6} w^{3} w^{6} w^{9} w^{13} w^{12} w^{8} 1 w^{4} 0$ & \texttt{ACAGACCATGGACGTTTAAA} \\
			\hline
			120 & $w^{11} w^{14} w^{3} w^{1} w^{13} w^{7} w^{4} w^{12} w^{11} w^{14}$ & \texttt{CTTCAGATTGGTTAGACTTC} \\
			\hline
			121 & $w^{4} w^{5} 1 w^{4} w^{13} w^{1} w^{10} w^{9} w^{14} w^{10}$ & \texttt{TACCTTTATGATGGCATCGG} \\
			\hline
			122 & $1 w^{10} w^{8} w^{7} w^{13} w^{5} 1 w^{2} w^{2} w^{11}$ & \texttt{TTGGCGGTTGCCTTGCGCCT} \\
			\hline
			123 & $w^{10} w^{2} w^{14} 0 w^{13} w^{9} w^{7} w^{8} 0 w^{5}$ & \texttt{GGGCTCAATGCAGTCGAACC} \\
			\hline
			124 & $w^{8} w^{8} w^{13} w^{3} w^{13} w^{11} w^{13} w^{7} w^{13} w^{12}$ & \texttt{CGCGTGAGTGCTTGGTTGGA} \\
			\hline
			125 & $w^{7} w^{1} w^{1} w^{8} w^{13} 1 w^{6} w^{10} w^{3} w^{8}$ & \texttt{GTATATCGTGTTACGGAGCG} \\
			\hline
			126 & $w^{14} w^{4} w^{5} w^{13} w^{13} w^{8} w^{9} w^{14} w^{8} w^{6}$ & \texttt{TCTACCTGTGCGCATCCGAC} \\
			\hline
			127 & $w^{12} 0 w^{9} w^{5} w^{13} w^{14} 0 w^{3} 1 w^{2}$ & \texttt{GAAACACCTGTCAAAGTTGC} \\
			\hline
			128 & $w^{13} 1 w^{11} w^{11} w^{13} w^{13} w^{5} w^{5} w^{6} w^{13}$ & \texttt{TGTTCTCTTGTGCCCCACTG} \\
			\hline
			129 & $0 w^{4} w^{3} w^{5} w^{6} w^{10} w^{9} w^{12} 1 w^{12}$ & \texttt{AATAAGCCACGGCAGATTGA} \\
			\hline
			130 & $w^{1} w^{1} w^{6} w^{11} w^{6} w^{4} w^{6} 1 w^{6} w^{5}$ & \texttt{ATATACCTACTAACTTACCC} \\
			\hline
			131 & $w^{2} 1 0 w^{8} w^{6} w^{12} w^{5} 0 w^{3} w^{11}$ & \texttt{GCTTAACGACGACCAAAGCT} \\
			\hline
			132 & $w^{5} 0 w^{2} w^{13} w^{6} w^{7} 0 w^{11} w^{8} w^{10}$ & \texttt{CCAAGCTGACGTAACTCGGG} \\
			\hline
			133 & $w^{3} w^{10} w^{7} 0 w^{6} w^{2} 1 w^{1} 0 w^{13}$ & \texttt{AGGGGTAAACGCTTATAATG} \\
			\hline
			134 & $w^{9} w^{5} w^{12} w^{3} w^{6} 0 w^{10} w^{6} w^{13} w^{2}$ & \texttt{CACCGAAGACAAGGACTGGC} \\
			\hline
			135 & $w^{6} w^{8} w^{4} w^{4} w^{6} w^{6} w^{13} w^{13} w^{14} w^{6}$ & \texttt{ACCGTATAACACTGTGTCAC} \\
			\hline
			136 & $w^{11} w^{2} w^{10} w^{7} w^{6} w^{3} w^{7} w^{4} w^{2} w^{8}$ & \texttt{CTGCGGGTACAGGTTAGCCG} \\
			\hline
			137 & $w^{4} w^{7} w^{11} w^{9} w^{6} 1 w^{12} w^{5} w^{4} w^{7}$ & \texttt{TAGTCTCAACTTGACCTAGT} \\
			\hline
			138 & $1 w^{9} w^{9} w^{1} w^{6} w^{8} w^{14} w^{3} w^{11} w^{1}$ & \texttt{TTCACAATACCGTCAGCTAT} \\
			\hline
			139 & $w^{10} w^{14} w^{5} w^{14} w^{6} w^{14} w^{4} w^{14} w^{9} w^{9}$ & \texttt{GGTCCCTCACTCTATCCACA} \\
			\hline
			140 & $w^{8} w^{3} w^{1} 1 w^{6} w^{13} w^{8} w^{10} w^{10} w^{4}$ & \texttt{CGAGATTTACTGCGGGGGTA} \\
			\hline
			141 & $w^{7} w^{12} w^{13} w^{6} w^{6} w^{1} w^{2} w^{7} w^{1} w^{14}$ & \texttt{GTGATGACACATGCGTATTC} \\
			\hline
			142 & $w^{14} w^{11} w^{14} w^{2} w^{6} w^{5} w^{1} w^{8} w^{12} 0$ & \texttt{TCCTTCGCACCCATCGGAAA} \\
			\hline
			143 & $w^{12} w^{13} w^{8} w^{12} w^{6} w^{9} w^{3} w^{2} w^{7} w^{3}$ & \texttt{GATGCGGAACCAAGGCGTAG} \\
			\hline
			144 & $w^{13} w^{6} 1 w^{10} w^{6} w^{11} w^{11} w^{9} w^{5} 1$ & \texttt{TGACTTGGACCTCTCACCTT} \\
			\hline
			145 & $0 1 w^{14} w^{1} w^{2} w^{6} w^{5} w^{8} w^{11} w^{8}$ & \texttt{AATTTCATGCACCCCGCTCG} \\
			\hline
			146 & $w^{1} 0 w^{13} w^{9} w^{2} w^{3} 0 w^{7} w^{4} w^{6}$ & \texttt{ATAATGCAGCAGAAGTTAAC} \\
			\hline
			147 & $w^{2} w^{4} 1 1 w^{2} w^{2} w^{9} w^{9} w^{10} w^{2}$ & \texttt{GCTATTTTGCGCCACAGGGC} \\
			\hline
			148 & $w^{5} w^{1} w^{8} w^{14} w^{2} 0 w^{6} w^{2} w^{9} w^{13}$ & \texttt{CCATCGTCGCAAACGCCATG} \\
			\hline
			149 & $w^{3} w^{8} w^{9} w^{2} w^{2} w^{12} w^{13} w^{3} w^{12} w^{10}$ & \texttt{AGCGCAGCGCGATGAGGAGG} \\
			\hline
			150 & $w^{9} w^{2} w^{11} w^{6} w^{2} w^{7} w^{7} w^{5} w^{1} w^{11}$ & \texttt{CAGCCTACGCGTGTCCATCT} \\
			\hline
			151 & $w^{6} w^{10} w^{1} w^{10} w^{2} w^{10} 1 w^{10} w^{5} w^{5}$ & \texttt{ACGGATGGGCGGTTGGCCCC} \\
			\hline
			152 & $w^{11} w^{5} w^{5} w^{12} w^{2} w^{4} w^{10} w^{14} w^{7} w^{12}$ & \texttt{CTCCCCGAGCTAGGTCGTGA} \\
			\hline
			153 & $w^{4} w^{14} w^{12} w^{11} w^{2} w^{9} w^{4} w^{6} w^{6} 1$ & \texttt{TATCGACTGCCATAACACTT} \\
			\hline
			154 & $1 w^{3} w^{7} w^{5} w^{2} w^{11} w^{8} w^{1} 1 w^{3}$ & \texttt{TTAGGTCCGCCTCGATTTAG} \\
			\hline
			155 & $w^{10} w^{7} w^{10} w^{13} w^{2} w^{1} w^{12} w^{4} w^{8} 0$ & \texttt{GGGTGGTGGCATGATACGAA} \\
			\hline
			156 & $w^{8} w^{9} w^{4} w^{8} w^{2} w^{5} w^{14} w^{13} w^{3} w^{14}$ & \texttt{CGCATACGGCCCTCTGAGTC} \\
			\hline
			157 & $w^{7} w^{13} w^{6} w^{3} w^{2} w^{14} w^{3} 1 w^{13} w^{4}$ & \texttt{GTTGACAGGCTCAGTTTGTA} \\
			\hline
			158 & $w^{14} w^{6} w^{3} 0 w^{2} w^{13} w^{11} w^{12} 0 w^{9}$ & \texttt{TCACAGAAGCTGCTGAAACA} \\
			\hline
			159 & $w^{12} w^{12} w^{2} w^{7} w^{2} 1 w^{2} w^{11} w^{2} w^{1}$ & \texttt{GAGAGCGTGCTTGCCTGCAT} \\
			\hline
			160 & $w^{13} w^{11} 0 w^{4} w^{2} w^{8} w^{1} 0 w^{14} w^{7}$ & \texttt{TGCTAATAGCCGATAATCGT} \\
			\hline
			161 & $0 w^{10} w^{9} w^{11} w^{12} w^{1} 1 w^{3} w^{6} w^{3}$ & \texttt{AAGGCACTGAATTTAGACAG} \\
			\hline
			162 & $w^{1} w^{5} w^{11} w^{5} w^{12} w^{5} w^{10} w^{5} 1 1$ & \texttt{ATCCCTCCGACCGGCCTTTT} \\
			\hline
			163 & $w^{2} w^{8} w^{1} w^{13} w^{12} w^{9} w^{13} w^{10} w^{8} w^{14}$ & \texttt{GCCGATTGGACATGGGCGTC} \\
			\hline
			164 & $w^{5} w^{2} w^{5} w^{8} w^{12} w^{11} w^{7} w^{14} w^{3} 0$ & \texttt{CCGCCCCGGACTGTTCAGAA} \\
			\hline
			165 & $w^{3} w^{4} w^{14} w^{3} w^{12} 1 w^{9} w^{8} w^{13} w^{9}$ & \texttt{AGTATCAGGATTCACGTGCA} \\
			\hline
			166 & $w^{9} w^{1} w^{13} 0 w^{12} w^{8} w^{6} w^{7} 0 w^{4}$ & \texttt{CAATTGAAGACGACGTAATA} \\
			\hline
			167 & $w^{6} 1 1 w^{7} w^{12} w^{14} w^{5} w^{9} w^{2} w^{7}$ & \texttt{ACTTTTGTGATCCCCAGCGT} \\
			\hline
			168 & $w^{11} 0 w^{8} w^{4} w^{12} w^{13} 0 w^{2} w^{14} w^{1}$ & \texttt{CTAACGTAGATGAAGCTCAT} \\
			\hline
			169 & $w^{4} w^{12} w^{6} w^{1} w^{12} w^{2} w^{2} 1 w^{11} w^{6}$ & \texttt{TAGAACATGAGCGCTTCTAC} \\
			\hline
			170 & $1 w^{11} w^{3} w^{9} w^{12} 0 w^{1} w^{12} w^{4} w^{8}$ & \texttt{TTCTAGCAGAAAATGATACG} \\
			\hline
			171 & $w^{10} w^{13} w^{2} 1 w^{12} w^{6} w^{3} w^{11} w^{10} w^{13}$ & \texttt{GGTGGCTTGAACAGCTGGTG} \\
			\hline
			172 & $w^{8} w^{6} 0 w^{14} w^{12} w^{3} w^{11} 0 w^{9} w^{2}$ & \texttt{CGACAATCGAAGCTAACAGC} \\
			\hline
			173 & $w^{7} w^{7} w^{12} w^{2} w^{12} w^{10} w^{12} w^{6} w^{12} w^{11}$ & \texttt{GTGTGAGCGAGGGAACGACT} \\
			\hline
			174 & $w^{14} w^{9} w^{7} w^{6} w^{12} w^{4} w^{14} w^{1} w^{1} w^{10}$ & \texttt{TCCAGTACGATATCATATGG} \\
			\hline
			175 & $w^{12} w^{14} w^{10} w^{10} w^{12} w^{12} w^{4} w^{4} w^{5} w^{12}$ & \texttt{GATCGGGGGAGATATACCGA} \\
			\hline
			176 & $w^{13} w^{3} w^{4} w^{12} w^{12} w^{7} w^{8} w^{13} w^{7} w^{5}$ & \texttt{TGAGTAGAGAGTCGTGGTCC} \\
			\hline
			177 & $0 w^{8} w^{7} w^{9} w^{10} w^{14} w^{13} w^{1} w^{4} w^{1}$ & \texttt{AACGGTCAGGTCTGATTAAT} \\
			\hline
			178 & $w^{1} w^{2} w^{12} w^{1} w^{10} w^{13} w^{7} w^{6} w^{11} w^{7}$ & \texttt{ATGCGAATGGTGGTACCTGT} \\
			\hline
			179 & $w^{2} w^{10} w^{4} w^{14} w^{10} 1 1 w^{13} w^{9} w^{4}$ & \texttt{GCGGTATCGGTTTTTGCATA} \\
			\hline
			180 & $w^{5} w^{5} w^{10} 1 w^{10} w^{8} w^{10} w^{4} w^{10} w^{9}$ & \texttt{CCCCGGTTGGCGGGTAGGCA} \\
			\hline
			181 & $w^{3} 1 w^{3} w^{6} w^{10} w^{9} w^{5} w^{12} w^{1} 0$ & \texttt{AGTTAGACGGCACCGAATAA} \\
			\hline
			182 & $w^{9} 0 w^{6} w^{2} w^{10} w^{11} 0 1 w^{12} w^{14}$ & \texttt{CAAAACGCGGCTAATTGATC} \\
			\hline
			183 & $w^{6} w^{4} 0 w^{12} w^{10} w^{1} w^{9} 0 w^{7} 1$ & \texttt{ACTAAAGAGGATCAAAGTTT} \\
			\hline
			184 & $w^{11} w^{1} w^{2} w^{10} w^{10} w^{5} w^{6} w^{11} w^{5} w^{3}$ & \texttt{CTATGCGGGGCCACCTCCAG} \\
			\hline
			185 & $w^{4} w^{13} w^{13} w^{5} w^{10} w^{12} w^{3} w^{7} 1 w^{5}$ & \texttt{TATGTGCCGGGAAGGTTTCC} \\
			\hline
			186 & $1 w^{6} w^{14} w^{11} w^{10} w^{7} w^{11} w^{8} w^{6} w^{12}$ & \texttt{TTACTCCTGGGTCTCGACGA} \\
			\hline
			187 & $w^{10} w^{12} w^{8} w^{8} w^{10} w^{10} w^{2} w^{2} w^{3} w^{10}$ & \texttt{GGGACGCGGGGGGCGCAGGG} \\
			\hline
			188 & $w^{8} w^{11} 1 w^{13} w^{10} w^{4} w^{1} w^{9} w^{8} w^{11}$ & \texttt{CGCTTTTGGGTAATCACGCT} \\
			\hline
			189 & $w^{7} w^{14} w^{11} 0 w^{10} w^{6} w^{4} w^{5} 0 w^{2}$ & \texttt{GTTCCTAAGGACTACCAAGC} \\
			\hline
			190 & $w^{14} w^{3} w^{9} w^{3} w^{10} w^{3} w^{8} w^{3} w^{13} w^{13}$ & \texttt{TCAGCAAGGGAGCGAGTGTG} \\
			\hline
			191 & $w^{12} w^{7} w^{5} w^{4} w^{10} w^{2} w^{12} w^{14} w^{14} w^{8}$ & \texttt{GAGTCCTAGGGCGATCTCCG} \\
			\hline
			192 & $w^{13} w^{9} w^{1} w^{7} w^{10} 0 w^{14} w^{10} w^{2} w^{6}$ & \texttt{TGCAATGTGGAATCGGGCAC} \\
			\hline
			193 & $0 w^{7} w^{6} w^{8} w^{9} w^{13} w^{12} 1 w^{3} 1$ & \texttt{AAGTACCGCATGGATTAGTT} \\
			\hline
			194 & $w^{1} w^{9} w^{3} w^{13} w^{9} w^{14} w^{14} w^{12} w^{8} w^{3}$ & \texttt{ATCAAGTGCATCTCGACGAG} \\
			\hline
			195 & $w^{2} w^{14} w^{2} w^{5} w^{9} w^{8} w^{4} w^{11} 1 0$ & \texttt{GCTCGCCCCACGTACTTTAA} \\
			\hline
			196 & $w^{5} w^{3} 0 w^{11} w^{9} 1 w^{8} 0 w^{6} w^{14}$ & \texttt{CCAGAACTCATTCGAAACTC} \\
			\hline
			197 & $w^{3} w^{12} w^{12} w^{4} w^{9} w^{11} w^{2} w^{6} w^{14} w^{4}$ & \texttt{AGGAGATACACTGCACTCTA} \\
			\hline
			198 & $w^{9} w^{11} w^{7} w^{7} w^{9} w^{9} w^{1} w^{1} w^{2} w^{9}$ & \texttt{CACTGTGTCACAATATGCCA} \\
			\hline
			199 & $w^{6} w^{13} w^{10} 0 w^{9} w^{5} w^{3} w^{4} 0 w^{1}$ & \texttt{ACTGGGAACACCAGTAAAAT} \\
			\hline
			200 & $w^{11} w^{6} w^{4} w^{3} w^{9} w^{1} w^{11} w^{13} w^{13} w^{7}$ & \texttt{CTACTAAGCAATCTTGTGGT} \\
			\hline
			201 & $w^{4} w^{4} w^{9} w^{14} w^{9} w^{7} w^{9} w^{3} w^{9} w^{8}$ & \texttt{TATACATCCAGTCAAGCACG} \\
			\hline
			202 & $1 w^{1} w^{11} 1 w^{9} w^{12} w^{6} w^{5} w^{10} w^{6}$ & \texttt{TTATCTTTCAGAACCCGGAC} \\
			\hline
			203 & $w^{10} 1 w^{1} w^{9} w^{9} w^{4} w^{5} w^{10} w^{4} w^{2}$ & \texttt{GGTTATCACATACCGGTAGC} \\
			\hline
			204 & $w^{8} 0 w^{5} w^{1} w^{9} w^{10} 0 w^{14} w^{11} w^{13}$ & \texttt{CGAACCATCAGGAATCCTTG} \\
			\hline
			205 & $w^{7} w^{10} w^{14} w^{12} w^{9} w^{3} 1 w^{8} w^{7} w^{10}$ & \texttt{GTGGTCGACAAGTTCGGTGG} \\
			\hline
			206 & $w^{14} w^{5} w^{13} w^{10} w^{9} w^{6} w^{10} w^{7} w^{5} w^{11}$ & \texttt{TCCCTGGGCAACGGGTCCCT} \\
			\hline
			207 & $w^{12} w^{8} 1 w^{6} w^{9} 0 w^{13} w^{9} w^{1} w^{5}$ & \texttt{GACGTTACCAAATGCAATCC} \\
			\hline
			208 & $w^{13} w^{2} w^{8} w^{2} w^{9} w^{2} w^{7} w^{2} w^{12} w^{12}$ & \texttt{TGGCCGGCCAGCGTGCGAGA} \\
			\hline
			209 & $0 w^{14} w^{13} 1 w^{1} w^{5} w^{4} w^{7} w^{10} w^{7}$ & \texttt{AATCTGTTATCCTAGTGGGT} \\
			\hline
			210 & $w^{1} w^{3} w^{14} w^{14} w^{1} w^{1} w^{8} w^{8} w^{9} w^{1}$ & \texttt{ATAGTCTCATATCGCGCAAT} \\
			\hline
			211 & $w^{2} w^{7} w^{8} w^{1} w^{1} w^{11} w^{12} w^{2} w^{11} w^{9}$ & \texttt{GCGTCGATATCTGAGCCTCA} \\
			\hline
			212 & $w^{5} w^{9} 1 w^{9} w^{1} w^{9} w^{14} w^{9} w^{4} w^{4}$ & \texttt{CCCATTCAATCATCCATATA} \\
			\hline
			213 & $w^{3} w^{13} w^{11} w^{10} w^{1} w^{8} w^{3} w^{5} w^{5} w^{14}$ & \texttt{AGTGCTGGATCGAGCCCCTC} \\
			\hline
			214 & $w^{9} w^{6} w^{9} w^{12} w^{1} 1 w^{11} w^{3} w^{7} 0$ & \texttt{CAACCAGAATTTCTAGGTAA} \\
			\hline
			215 & $w^{6} w^{12} w^{5} w^{2} w^{1} w^{13} w^{2} w^{14} w^{12} w^{3}$ & \texttt{ACGACCGCATTGGCTCGAAG} \\
			\hline
			216 & $w^{11} w^{11} w^{1} w^{6} w^{1} w^{14} w^{1} w^{10} w^{1} 1$ & \texttt{CTCTATACATTCATGGATTT} \\
			\hline
			217 & $w^{4} 1 w^{7} w^{13} w^{1} 0 w^{5} w^{1} w^{8} w^{12}$ & \texttt{TATTGTTGATAACCATCGGA} \\
			\hline
			218 & $1 0 w^{12} w^{8} w^{1} w^{2} 0 w^{6} w^{3} w^{5}$ & \texttt{TTAAGACGATGCAAACAGCC} \\
			\hline
			219 & $w^{10} w^{4} w^{4} w^{11} w^{1} w^{3} w^{9} w^{13} w^{6} w^{11}$ & \texttt{GGTATACTATAGCATGACCT} \\
			\hline
			220 & $w^{8} w^{1} w^{10} w^{5} w^{1} w^{6} w^{6} w^{4} 1 w^{10}$ & \texttt{CGATGGCCATACACTATTGG} \\
			\hline
			221 & $w^{7} w^{8} w^{3} w^{7} w^{1} w^{4} w^{13} w^{12} w^{2} w^{13}$ & \texttt{GTCGAGGTATTATGGAGCTG} \\
			\hline
			222 & $w^{14} w^{2} w^{6} w^{4} w^{1} w^{10} w^{7} 1 w^{14} w^{2}$ & \texttt{TCGCACTAATGGGTTTTCGC} \\
			\hline
			223 & $w^{12} w^{10} 0 w^{3} w^{1} w^{7} 1 0 w^{13} w^{6}$ & \texttt{GAGGAAAGATGTTTAATGAC} \\
			\hline
			224 & $w^{13} w^{5} w^{2} 0 w^{1} w^{12} w^{10} w^{11} 0 w^{8}$ & \texttt{TGCCGCAAATGAGGCTAACG} \\
			\hline
			225 & $0 w^{12} w^{11} w^{13} w^{14} w^{3} w^{2} w^{5} w^{8} w^{5}$ & \texttt{AAGACTTGTCAGGCCCCGCC} \\
			\hline
			226 & $w^{1} w^{11} w^{9} w^{8} w^{14} w^{6} w^{1} w^{3} w^{3} w^{12}$ & \texttt{ATCTCACGTCACATAGAGGA} \\
			\hline
			227 & $w^{2} w^{13} w^{5} w^{11} w^{14} 0 w^{3} w^{14} w^{6} w^{10}$ & \texttt{GCTGCCCTTCAAAGTCACGG} \\
			\hline
			228 & $w^{5} w^{6} w^{1} w^{5} w^{14} w^{2} w^{11} w^{10} 1 w^{11}$ & \texttt{CCACATCCTCGCCTGGTTCT} \\
			\hline
			229 & $w^{3} w^{7} w^{13} w^{7} w^{14} w^{7} w^{12} w^{7} w^{2} w^{2}$ & \texttt{AGGTTGGTTCGTGAGTGCGC} \\
			\hline
			230 & $w^{9} w^{9} w^{14} w^{4} w^{14} w^{12} w^{14} w^{8} w^{14} w^{13}$ & \texttt{CACATCTATCGATCCGTCTG} \\
			\hline
			231 & $w^{6} w^{14} w^{8} w^{3} w^{14} w^{4} w^{4} w^{2} w^{13} w^{8}$ & \texttt{ACTCCGAGTCTATAGCTGCG} \\
			\hline
			232 & $w^{11} w^{3} 1 0 w^{14} w^{10} w^{8} w^{9} 0 w^{6}$ & \texttt{CTAGTTAATCGGCGCAAAAC} \\
			\hline
			233 & $w^{4} w^{10} w^{3} 1 w^{14} w^{11} 1 w^{12} w^{10} w^{1}$ & \texttt{TAGGAGTTTCCTTTGAGGAT} \\
			\hline
			234 & $1 w^{5} w^{6} w^{14} w^{14} w^{9} w^{10} 1 w^{9} w^{7}$ & \texttt{TTCCACTCTCCAGGTTCAGT} \\
			\hline
			235 & $w^{10} w^{8} 0 w^{1} w^{14} w^{5} w^{13} 0 w^{11} w^{4}$ & \texttt{GGCGAAATTCCCTGAACTTA} \\
			\hline
			236 & $w^{8} w^{2} w^{2} w^{9} w^{14} w^{1} w^{7} w^{11} w^{4} w^{9}$ & \texttt{CGGCGCCATCATGTCTTACA} \\
			\hline
			237 & $w^{7} w^{4} w^{7} w^{10} w^{14} w^{13} w^{9} w^{1} w^{5} 0$ & \texttt{GTTAGTGGTCTGCAATCCAA} \\
			\hline
			238 & $w^{14} w^{1} w^{12} w^{12} w^{14} w^{14} w^{6} w^{6} w^{7} w^{14}$ & \texttt{TCATGAGATCTCACACGTTC} \\
			\hline
			239 & $w^{12} 1 w^{4} w^{2} w^{14} w^{8} w^{5} w^{13} w^{12} 1$ & \texttt{GATTTAGCTCCGCCTGGATT} \\
			\hline
			240 & $w^{13} 0 w^{10} w^{6} w^{14} 1 0 w^{4} w^{1} w^{3}$ & \texttt{TGAAGGACTCTTAATAATAG} \\
			\hline
			241 & $0 w^{13} w^{12} w^{14} 1 w^{4} w^{3} w^{6} w^{9} w^{6}$ & \texttt{AATGGATCTTTAAGACCAAC} \\
			\hline
			242 & $w^{1} w^{6} w^{7} 1 1 w^{10} w^{11} w^{1} w^{10} w^{8}$ & \texttt{ATACGTTTTTGGCTATGGCG} \\
			\hline
			243 & $w^{2} w^{12} w^{10} w^{9} 1 w^{7} w^{2} w^{4} w^{4} w^{13}$ & \texttt{GCGAGGCATTGTGCTATATG} \\
			\hline
			244 & $w^{5} w^{11} w^{4} w^{1} 1 w^{12} w^{1} w^{13} w^{11} w^{2}$ & \texttt{CCCTTAATTTGAATTGCTGC} \\
			\hline
			245 & $w^{3} w^{14} w^{6} w^{12} 1 0 w^{4} 1 w^{7} w^{11}$ & \texttt{AGTCACGATTAATATTGTCT} \\
			\hline
			246 & $w^{9} w^{3} w^{3} w^{10} 1 w^{2} w^{8} w^{12} w^{5} w^{10}$ & \texttt{CAAGAGGGTTGCCGGACCGG} \\
			\hline
			247 & $w^{6} w^{7} w^{2} w^{6} 1 w^{3} w^{12} w^{11} w^{1} w^{12}$ & \texttt{ACGTGCACTTAGGACTATGA} \\
			\hline
			248 & $w^{11} w^{9} 0 w^{2} 1 w^{6} w^{14} 0 w^{12} w^{5}$ & \texttt{CTCAAAGCTTACTCAAGACC} \\
			\hline
			249 & $w^{4} w^{8} w^{14} w^{8} 1 w^{8} w^{13} w^{8} w^{3} w^{3}$ & \texttt{TACGTCCGTTCGTGCGAGAG} \\
			\hline
			250 & $1 w^{2} w^{13} w^{13} 1 1 w^{7} w^{7} w^{8} 1$ & \texttt{TTGCTGTGTTTTGTGTCGTT} \\
			\hline
			251 & $w^{10} w^{10} 1 w^{5} 1 w^{13} 1 w^{9} 1 w^{14}$ & \texttt{GGGGTTCCTTTGTTCATTTC} \\
			\hline
			252 & $w^{8} w^{5} w^{8} w^{11} 1 w^{14} w^{10} w^{2} w^{6} 0$ & \texttt{CGCCCGCTTTTCGGGCACAA} \\
			\hline
			253 & $w^{7} 1 w^{9} w^{4} 1 w^{5} w^{5} w^{3} w^{14} w^{9}$ & \texttt{GTTTCATATTCCCCAGTCCA} \\
			\hline
			254 & $w^{14} 0 w^{11} w^{7} 1 w^{1} 0 w^{5} w^{2} w^{4}$ & \texttt{TCAACTGTTTATAACCGCTA} \\
			\hline
			255 & $w^{12} w^{4} w^{1} 0 1 w^{11} w^{9} w^{10} 0 w^{7}$ & \texttt{GATAATAATTCTCAGGAAGT} \\
			\hline
			256 & $w^{13} w^{1} w^{5} w^{3} 1 w^{9} w^{6} w^{14} w^{13} w^{1}$ & \texttt{TGATCCAGTTCAACTCTGAT} \\
			\hline
		\end{longtable}
	\end{exl}
	\section{ Self-dual and LCD codes constructions}
	\begin{propo}\cite[Theorem 4.5]{Chaturvedi}\label{self dual reversible}  Let $n$ be a positive even integer. Let $F_q$ be a finite field of order $q$ and $M$ be a $k\times n$ matrix over $F_q$ such that $rank(M)=k$ and $k=\frac{n}{2}$. Let $D$ be a $k\times k$ involutory matrix over $F_q$. Then the linear code generated by $[M\mid M_D']$ is a self dual reversible linear code if and only if the matrix $DMM^T$ is skew symmetric, where $M^T$ is the transpose of $M.$
			
	\end{propo}
	\begin{lem}\cite[Corollary 4.6]{Chaturvedi}\label{self duality preserved}
Let $n$ be a positive even integer. Let $F_q$ be a finite field of order $q$ and $M$ be a $k\times n$ matrix over $F_q$ such that $rank(M)=k$ and $k=\frac{n}{2}$. Let $D$ be a $k\times k$ involutory matrix over $F_q$.
If the linear code generated by $M$ is self dual then the linear code generated by $[M\mid M_D']$ is a self dual reversible linear code.
	\end{lem}
	
	\begin{lem}\label{self dual cyclic}
		Let $\mathbb{F}_q$ be a finite field and $n$ be a natural number. Let $g(x)$ be a monic divisor of $x^n-1$ in $\mathbb{F}_q[x]$ and $h(x)=(x^n-1)/g(x)$. Let $g^*(x)$ and $h^*(x)$ denote the reciprocal polynomials of $g(x)$ and $h(x)$, respectively. Then the cyclic code $C=\langle g(x)\rangle$ is self-dual if and only if $g(x)g^*(x)=c(1-x^n)$ for some $c\in\mathbb{F}_q^*$.
	\end{lem}
	\begin{proof}
		Since $g(x)$ divides $x^n-1$, the dual code $C^\perp$ is the cyclic code generated by $h^*(x)$, that is, $C^\perp=\langle h^*(x)\rangle$. Hence $C$ is self-dual if and only if $\langle g(x)\rangle=\langle h^*(x)\rangle$. Since two nonzero polynomials generate the same ideal in $\mathbb{F}_q[x]$ if and only if they are nonzero scalar multiples of each other, this is equivalent to $g(x)=c h^*(x)$ for some $c\in\mathbb{F}_q^*$. By definition of reciprocal polynomial, $g(x)=cx^{n-deg(g)}h(x^{-1})$. Therefore, $g(x)=cx^{n-deg(g)}\frac{x^{-n}-1}{g(x^{-1})}.$ Solving further, $g(x)=c\frac{x^{-deg(g)}-x^{n-deg(g)}}{g(x^{-1})}.$ Hence, $g(x)=c\frac{1-x^n}{g*(x)}.$ Thus, $g(x)g^*(x)=c(1-x^n).$
	\end{proof}
	\begin{thm}\label{reversible self dual}
		Let $\mathbb{F}_q$ be a finite field, and $n$ be a positive even integer.  Let $g(x)=z_0+z_1x+\cdots+z_sx^s\in\mathbb{F}_q[x]$ such that $s=\frac{n}{2}$. Let $m=n-s$. Let $g^*(x)$ be the reciprocal polynomial of $g(x)$ such that $g(x)g^*(x)=c(x^n-1)$ for some $c\in\mathbb{F}_q^*$. Then $\langle T_g^n\mid T_{D,g}^n\rangle$ is a reversible self-dual linear code.
	\end{thm}
	\begin{proof}
		By Theorem \ref{linear theorem},  $\langle T_g^n\mid T_{D,g}^n\rangle$ is a reversible linear code. Since $g(x)g^*(x)=c(x^n-1)$, therefore, by Lemma \ref{self dual cyclic}, the cyclic code generated by $g(x)$, that is, $C=\langle g(x)\rangle$ is self dual. It is easy to observe that the cyclic code generated by $g(x)$ is same as the linear code $<T_g^n>.$ Hence, $<T_g^n>$ is a self dual linear code. If we take $M=G_{T_g^n}$, then by \ref{MainLemma}, the double reflected matrix $M_D'$ is equal to $G_{T_D,g^n}$. Moreover, since $s=\frac{n}{2}$ and $G_{T_g^n}$ is an $(n-s)\times n$ matrix, therefore, the number of rows in $G_{T_g^n},$ that is, $m=n-s=n-\frac{n}{2}=\frac{n}{2}.$ Hence, by  Lemma \ref{self duality preserved}, $\langle T_g^n\mid T_{D,g}^n\rangle$ is a self-dual linear code. Therefore, $\langle T_g^n\mid T_{D,g}^n\rangle$ is a reversible self-dual linear code. 
	\end{proof}
	\begin{lem}\cite[Proposition 1]{Massey}\label{characterisation of lcd codes} If $G$ is a generator matrix for the (n, k) linear code $C$, then $C$ is an LCD code if and only if the $k\times k$ matrix $GG^T$ is nonsingular.
	\end{lem}
	\begin{lem}\cite[Theorem 1.3.1]{Golub}\label{block product} If
		\[
		A=
		\begin{array}{c}
			\left[
			\begin{array}{ccc}
				A_{11} & \cdots & A_{1s}\\
				\vdots &        & \vdots\\
				A_{q1} & \cdots & A_{qs}
			\end{array}
			\right]
			\\[-2mm]
			\begin{array}{ccc}
				p_1 && p_s
			\end{array}
		\end{array}
		\begin{array}{c}
			m_1\\
			\\
			m_q
		\end{array},
		\qquad
		B=
		\begin{array}{c}
			\left[
			\begin{array}{ccc}
				B_{11} & \cdots & B_{1r}\\
				\vdots &        & \vdots\\
				B_{s1} & \cdots & B_{sr}
			\end{array}
			\right]
			\\[-2mm]
			\begin{array}{ccc}
				n_1 && n_r
			\end{array}
		\end{array}
		\begin{array}{c}
			p_1\\
			\\
			p_s
		\end{array},
		\]
		and we partition the product \(C=AB\) as follows,
		\[
		C=
		\begin{array}{c}
			\left[
			\begin{array}{ccc}
				C_{11} & \cdots & C_{1r}\\
				\vdots &        & \vdots\\
				C_{q1} & \cdots & C_{qr}
			\end{array}
			\right]
			\\[-2mm]
			\begin{array}{ccc}
				n_1 && n_r
			\end{array}
		\end{array}
		\begin{array}{c}
			m_1\\
			\\
			m_q
		\end{array},
		\]
		then for \(\alpha=1\!:\!q\) and \(\beta=1\!:\!r\) we have
		\[
		C_{\alpha\beta}
		=
		\sum_{\gamma=1}^{s}
		A_{\alpha\gamma}B_{\gamma\beta}.
		\]
	\end{lem}
	\begin{propo}\label{lcd characterisation} Let $F_q$ be a finite field of order $q$ and $M$ be a $k\times k$ matrix over $F_q$ such that $rank(M)=k.$ Let $MM^T=A$ and $D$ be a $k\times k$ involutory matrix over $R$. Then the linear code generated by $[M|M_D']$ is a linear complementary dual code if and only if $det(A+DAD^T)\neq 0$.     	 	
	\end{propo}
	\begin{proof}
		Consider $G=[M|MD']=[M|DMJ_n].$ It is easy to verify that for two matrices $A$ and $B$ of order $k\times n$ over $F_q$, $[A|B]^T=\begin{bmatrix} 
			A^T\\
			B^T
		\end{bmatrix}.$ Hence, $$GG^T=[M|DMJ_n][M|DMJ_n]^T=[M|DMJ_n]\begin{bmatrix}
			M^T\\
			J_n^TM^TD^T
		\end{bmatrix}.$$ Therefore, using Lemma \ref{block product}, $GG^T=MM^T+DMJ_nJ_n^TM^TD^T.$ Since $J_n$ is symmetric and involutory, $J_nJ_n^T=I_d.$ Hence, $GG^T=MM^T+DMM^TD^T=A+DAD^T.$ By Lemma \ref{characterisation of lcd codes}, the linear code generated by $G$ is LCD if and only if $GG^T=A+DAD^T$ is non singular, that is, $det(A+DAD^T)\neq 0.$
	\end{proof}
	\begin{coro}
		Let $F_q$ be a finite field of order $q$ and $M$ be a $k\times k$ matrix over $F_q$ such that $rank(M)=k.$ Let $MM^T=A$ and $D$ be a symmetric $k\times k$ involutory matrix over $F_q$ such that $D$ commutes with $A$. Then the linear code generated by $[M|M_D']$ is a reversible linear complementary dual code if and only if the linear code generated by $M$ is a linear complementary dual code.
	\end{coro}
	\begin{proof}
		By Lemma \ref{linear2} the linear code generated by $[M|M_D']$ is always reversible. By Theorem \ref{lcd characterisation}, the linear code generated by $[M|M_D']$ is LCD if and only if $det(A+DAD^T)\neq 0.$ Since $D$ commutes with $A$ and $D$ is a symmetric involutory matrix, therefore $DD^T$ is identity. Hence, the linear code generated by $[M|M_D']$ is LCD if and only if $det(A)\neq 0.$ Equivalently, $A=MM^T$ is non singular. By Lemma \ref{characterisation of lcd codes}, this is equivalent to the fact that the linear code generated by $M$ is LCD code. 
	\end{proof}
	
	\section{$(n_1,c_1|n_2,c_2)-$right polycyclic codes}
	
	\begin{dfn}\cite[Definition 1]{matsuoka}\label{polycyclic definition} Let $C$ be a linear code of length $n$ over $R$. $C$ is a polycyclic code induced by $c$ if there exists a vector $c=(c_0,c_1,\ldots,c_{n-1})\in R^n$ such that for every $(a_0,a_1,\ldots,a_{n-1})$$\in C$, $(0,a_0,a_1,\ldots,a_{n-2})+$$ a_{n-1}(c_0,c_1,\ldots,c_{n-1})\in C.$ In this case we call $c$ an associated vector of $C$.
	\end{dfn}
	
	\begin{dfn}\cite[Definition 1]{Suxsena}\label{polyshift definition}
		Let $(y_0,y_1,...,y_{n-1})$ and $\mathbf{a}=(a_0,a_1,\ldots,a_{n-1})$ be two vectors in $\mathbb{F}_q^n$ with $a_0\neq 0$
		The right polyshift operator		$P_r^{\mathbf{a}}:\mathbb{F}_q^n\to\mathbb{F}_q^n$ is given by
		$$(y_0,y_1,\ldots,y_{n-1}) \rightarrow		y_{n-1}\mathbf{a}+(0,y_0,\ldots,y_{n-2}).$$
	\end{dfn}
	\begin{rmk} 
	Here, we give a notion of block wise polycyclicity for two blocks over any arbitrary finite commutative ring with unity. We denote the polyshift operator $P_r^{\mathbf{a}}$ (Definition \ref{polyshift definition}) by $\tau_{a}$ and call it an {\em $a$-polycylic shift operator}. Let $R$ be a finite commutative ring with unity, and $n_1,n_2$ be positive integers. Let
	$c_1=(c_{10},c_{11},\ldots,c_{1,n_1-1})\in R^{n_1}$	and $c_2=(c_{20},c_{21},\ldots,c_{2,n_2-1})\in R^{n_2}$
	be fixed vectors, called the \emph{associated vectors}.	For $u=(u_0,u_1,\ldots,u_{n_1-1})\in R^{n_1}$, we define $c_1$-polycyclic shift operator by $\tau_{c_1}(u)=(0,u_0,u_1,\ldots,u_{n_1-2})+u_{n_1-1}c_1.$
	Similarly, for $v=(v_0,v_1,\ldots,v_{n_2-1})\in R^{n_2}$, we define the $c_2$-polycyclic shift operator by $\tau_{c_2}(v)=(0,v_0,v_1,\ldots,v_{n_2-2})+v_{n_2-1}c_2$. In the following, we introduce the notion of $(n_1,c_1|n_2,c_2)-$right polycyclic code over $R$.
\end{rmk}
	\begin{dfn}\label{(n1,c1)}
		Let $R$ be any arbitrary finite commutative ring with unity and $C$ be an $R$-submodule of $R^{n_1}\times R^{n_2}$. Then $C$ is called an \emph{$(n_1,c_1|n_2,c_2)-$right polycyclic code} over $R$ if
		$(u,v)\in C$ implies that $\bigl(\tau_{c_1}(u),\tau_{c_2}(v)\bigr)\in C$. In other words, we say that $C$ is simultaneously invariant under the  $c_1$-polycyclic shift operator to the first component and the $c_2$-polycyclic shift operator to the second component.
	\end{dfn}
	
	Recall from \cite{matsuoka}, as cyclic codes, polycyclic codes may be understood in terms of ideals
	in quotient rings of polynomial rings. Given
	$c=(c_0,c_1,\ldots,c_{n-1})\in R^n$, if we let
	$f(X)=X^n-c(X)$, where
	$
	c(X)=c_{n-1}X^{n-1}+\cdots+c_1X+c_0,
	$
	then the $R$-module homomorphism
	$
	\rho:R^n\longrightarrow R[X]/(f(X))
	$
	sending the vector $a=(a_0,a_1,\ldots,a_{n-1})$ to the equivalence
	class of the polynomial
	$
	a_{n-1}X^{n-1}+\cdots+a_1X+a_0,
	$
	allows us to identify the polycyclic codes induced by $c$ with the
	ideal of $R[X]/(f(X))$.
	
	In the following remark we give an algebraic characterization of a \linebreak $(n_1,c_1|n_2,c_2)-$right polycyclic code.
	
	\begin{rmk}
		Definition \ref{(n1,c1)} admits a natural polynomial representation. Associate each vector
		$u=(u_0,u_1,\ldots,u_{n_1-1})\in R^{n_1}$
		with the polynomial
		$u(x)=u_0+u_1x+\cdots+u_{n_1-1}x^{n_1-1}$,
		and each vector
		$v=(v_0,v_1,\ldots,v_{n_2-1})\in R^{n_2}$
		with
		$v(x)=v_0+v_1x+\cdots+v_{n_2-1}x^{n_2-1}$.
		Similarly, associate the vectors $c_1$ and $c_2$ with the polynomials
		$c_1(x)=c_{10}+c_{11}x+\cdots+c_{1(n_1-1)}x^{n_1-1}$
		and
		$c_2(x)=c_{20}+c_{21}x+\cdots+c_{2(n_2-1)}x^{n_2-1}$.
		
		Under these identifications, each $c_i$-polycyclic shift corresponds to multiplication by $x$ modulo $x^{n_i}-c_i(x)$. More precisely,
		$
		\tau_{c_i}(u)$ is equivalent to $xu(x)\pmod{x^{n_i}-c_i(x)}, i=1,2.$
		Consequently, an $(n_1,c_1|n_2,c_2)-$right polycyclic code can equivalently be viewed as an $R[x]$-submodule of
		$
		\frac{R[x]}{\langle x^{n_1}-c_1(x)\rangle}
		\times
		\frac{R[x]}{\langle x^{n_2}-c_2(x)\rangle}
		$
		that is invariant under simultaneous multiplications by $x$ in the two components.
	\end{rmk}
	\begin{exl}\label{partwise poly need not be poly}
		Let $R=\mathbb{F}_2, n_1=n_2=3,$ and $c_1=c_2=(1,0,0).$ Thus, both component shifts are the ordinary cyclic shift. Consider the code $C=\{(u\mid u):u\in\mathbb{F}_2^3\}\subseteq \mathbb{F}_2^3\times\mathbb{F}_2^3.$ Explicitly,
		$$C=\{000000,\;001001,\;010010,\;011011,\;100100,\;101101,\;110110,\;111111\}.$$	
		For any $(u\mid u)\in C$, applying the cyclic shift independently to the
		two components gives $(\tau_{c_1}(u)\mid \tau_{c_2}(u))
		=(\tau(u)\mid \tau(u)),$ where $\tau(a_1,a_2,a_3)=(a_3,a_1,a_2)$ is the ordinary cyclic shift on $\mathbb{F}_2^3$. Since the independent cyclic shifts again belong to $C,$ therefore, $C$ is a $(3,c_1|3,c_2)-$right polycyclic code.
		
		Now regard $C$ as a code of length $6$ over $\mathbb{F}_2$. Suppose, for a contradiction, that $C$ is an ordinary polycyclic code with respect to	some associated vector $c=(c_0,c_1,c_2,c_3,c_4,c_5)\in\mathbb{F}_2^6.$ Consider the codeword $x=010010\in C.$ Then the polycyclic shift of $x$ is given by $\tau_c(x)=000101$ which does not belong to $C.$
		This contradicts the assumption that $C$ is invariant under an
		ordinary polycyclic shift. Hence $C$ is not an ordinary polycyclic code of length $6$.
	\end{exl}
	
	In the following, we highlight the non trivial nature of the \linebreak $(n_1,c_1|n_2,c_2)-$right polycylic code. By merely concatenating two generator matrices which independently generate polycylic codes, we may not get a generator matrix for $(n_1,c_1|n_2,c_2)-$right polycylic code.
	\begin{rmk}
		Let $R$ be a finite commutative ring with unity. Let $A$ and $B$ be matrices of order $k\times n_1$ and $k\times n_2$ over $R$ such that the codes generated over $R$ by $A$ and $B$ are polycylic over $R$ with respect to associated vectors $c_1$ and $c_2$, respectively. If we use the definition of concatenation as given in Definition \ref{concatenation}, then it is not necessary that the linear code over $R$ generated by $[A|B]$ is an $(n_1,c_1|n_2,c_2)-$right polycyclic code. This is illustrated by the following example.
	\end{rmk}
	\begin{exl}\label{concatenation of poly not partwise poly}
		Let $R=Z_2$ and $n_1=n_2=2$. Consider two generator matrices $A$ and $B$, both Identity matrices of order $2$ over $R$. Since they are of rank $2$ each, their row spaces and hence the linear codes generated by them span the whole ambient space $Z_2^2$. Hence $A$ and $B$ generate a polycylic code with respect to any chosen associated vectors $c_1$ and $c_2$ in $Z_2^2.$ In particular, for $c_1=(1,0)$ and $c_2=(0,1)$, $A$ generates a $c_1$ polycyclic code and $B$ generates a $c_2$ polycyclic code over $Z_2.$ Consider the concatenation of the matrices $A$ and $B$, that is, $G=[A|B].$ Then $G=\begin{bmatrix}
			1 & 0 & 1 & 0\\
			0 & 1 & 0 & 1
		\end{bmatrix}$. Computing explicitly the linear code $C$ spanned by $G$, we find $C=\{(0,0,0,0), (0,1,1,0), (1,0,1,0), (1,1,1,1)\}$. For the codeword $(0,1,0,1)$ in $C$, the $(c_1,c_2)$- polycyclic shift is calculated to be $(1,0,0,1)$ which is not in $C$. Hence $C$ is not a $(2,c_1|2,c_2)-$right polycyclic code. 
	\end{exl}
	
	Thus a natural question arises that "When does the concatenation of two generator matrices of polycyclic codes again a generator matrix for a $(n_1,c_1|n_2,c_2)-$right polycyclic code?"
	 In  the following, we provide a sufficient condition for existence of such codes.
	\begin{propo}\label{5} Let $A$ and $B$ be any two matrices of order $r\times n_1$ and $r\times n_2$ over a finite commutative ring with unity $R$, such that they generate polycyclic codes with respect to associated vectors $c_1$ and $c_2$ respectively. Further, let for all $x\in R^r,$ the $c_1$ polycyclic shift of the codeword $xA$ and the $c_2$ polycylic shift of $xB$ are given by $xMA$ and $xMb$ for some $M\in R^{r\times r}.$ Then $[A|B]$ generates an $(n_1,c_1|n_2,c_2)-$right polycyclic code.
	\end{propo}
	
	\begin{proof}
		Let $(u\mid v)\in C$. Then there exists $x\in R^r$ such that $(u\mid v)=x[A\mid B]=(xA\mid xB)$. Since $\tau_{\mathbf{c_1}}(A)=MA$, the first component satisfies $\tau_{\mathbf{c_1}}(u)=\tau_{\mathbf{c_1}}(xA)=xMA$. Similarly, since $\tau_{\mathbf{c_2}}(B)=MB$, we have $\tau_{\mathbf{c_2}}(v)=\tau_{\mathbf{c_2}}(xB)(B)=xMB$. Hence $(\tau_{\mathbf{c_1}}(u)\mid\tau_{\mathbf{c_2}}(v))=(xMA\mid xMB)=(xM)[A\mid B]$. Since $xM\in R^r$, we have $(xM)[A\mid B]\in\operatorname{Row}[A\mid B]=C$. Therefore $(u\mid v)\in C$ implies $(\tau_{\mathbf{c_1}}(u)\mid\tau_{\mathbf{c_2}}(v))\in C$ and hence $C$ is an $(n_1,c_1|n_2,c_2)-$right polycyclic code.
	\end{proof}
	
	\begin{qn}\label{open} Is the converse of Proposition \ref{5} true?
		\end{qn}
	\begin{lem}\label{polycyclic shift in terms of Sm}
		Let $R$ be a finite commutative ring with unity, let $G\in R^{m\times n}$ have rows $g_1,g_2,\ldots,g_m$, and let $u=(u_1,u_2,\ldots,u_m)\in R^m$. Let $\tau_{c}:R^n\to R^n$ be a polycyclic shift satisfying
		$\tau_{c}(g_i)=g_{i+1}$ for $1\le i<m$ and $\tau_{c}(g_m)=-g_1$. Then for
		$S_m=\begin{pmatrix}
			0&1&0&\cdots&0\\
			0&0&1&\cdots&0\\
			\vdots&\vdots&\ddots&\ddots&\vdots\\
			0&0&\cdots&0&1\\
			-1&0&\cdots&0&0
		\end{pmatrix}, \tau_{c}(uG)=uS_mG.$
	\end{lem}
	
	\begin{proof}
		Writing $G$ in terms of its rows, we have
		$uG=\sum_{i=1}^{m}u_i g_i$. By the $R$-linearity of $\tau_{c}$,
		\[
		\begin{aligned}
			\tau_{c}(uG)
			&=\tau_{c}\left(\sum_{i=1}^{m}u_i g_i\right)\\
			&=\sum_{i=1}^{m}u_i\tau_{c}(g_i)\\
			&=\sum_{i=1}^{m-1}u_i g_{i+1}-u_mg_1\\
			&=(-u_m)g_1+\sum_{i=1}^{m-1}u_i g_{i+1}.
		\end{aligned}
		\]
		On the other hand, by the definition of $S_m$,
		$uS_m=(-u_m,u_1,u_2,\ldots,u_{m-1})$. Therefore,
		\[
		\begin{aligned}
			uS_mG
			&=(-u_m,u_1,\ldots,u_{m-1})
			\begin{pmatrix}
				g_1\\
				g_2\\
				\vdots\\
				g_m
			\end{pmatrix}\\
			&=(-u_m)g_1+\sum_{i=1}^{m-1}u_i g_{i+1}\\
			&=\tau_{c}(uG).
		\end{aligned}
		\]
		Hence $\tau_{c}(uG)=uS_mG$.
	\end{proof}
	\begin{lem}\label{Sm lemma}
		Let $R$ be a commutative ring with unity. Let
		\[
		S_m=
		\begin{pmatrix}
			0&1&0&\cdots&0\\
			0&0&1&\cdots&0\\
			\vdots&\vdots&\ddots&\ddots&\vdots\\
			0&0&\cdots&0&1\\
			-1&0&\cdots&0&0
		\end{pmatrix}
		\]
		and let $J_m$ be the $m\times m$ matrix with ones on anti diagonal and zeros elsewhere. Then
		$J_mS_mJ_m=S_m^{-1}$.
	\end{lem}
	
	\begin{proof}
		Let $v=(v_1,v_2,\ldots,v_m)\in R^m$ be arbitrary. Since $J_m$ reverses the coordinates,
		\[
		vJ_m=(v_m,v_{m-1},\ldots,v_1).
		\]
		Multiplication by $S_m$ gives
		\[
		vJ_mS_m=(-v_1,v_m,v_{m-1},\ldots,v_2).
		\]
		Applying $J_m$ once again, we obtain
		\[
		vJ_mS_mJ_m=(v_2,v_3,\ldots,v_m,-v_1).
		\]
		
		On the other hand, since
		\[
		S_m^{-1}=
		\begin{pmatrix}
			0&0&\cdots&0&-1\\
			1&0&\cdots&0&0\\
			0&1&\cdots&0&0\\
			\vdots&\ddots&\ddots&\vdots&\vdots\\
			0&\cdots&0&1&0
		\end{pmatrix},
		\]
		we have
		\[
		vS_m^{-1}=(v_2,v_3,\ldots,v_m,-v_1).
		\]
		Therefore, $vJ_mS_mJ_m=vS_m^{-1}$ for every $v\in R^m$. Hence,
		\[
		J_mS_mJ_m=S_m^{-1}.
		\]
	\end{proof}
	\begin{propo}\label{partwise polycylic}
		Let $R$ be a finite commutative ring with unity and $g(x)=1+z_1x+\cdots+z_{s-1}x^{s-1}+x^s\in R[x]$, be a monic polynomial of degree $s<n.$ Let $n=s+m$ with $m\geq 1$. Let $S_m$ be the $m\times m$ matrix defined by $S_m=\begin{bmatrix}0&1&0&\cdots&0\\
			0&0&1&\cdots&0\\ 
			\vdots&\vdots&\ddots&\vdots&1\\
			-1&0&\cdots&0&0
		\end{bmatrix}.$
		For $k=0,1,\ldots,m-1$, let $D_k=S_m^kJ_m$. Let $c_1(x)=x^n-g(x)(x^m+1)$ and $c_2(x)=x^n-g^*(x)(x^m+1)$ and $\mathbf{c}_1$ and $\mathbf{c}_2$ be the vectors whose entries are the coefficients of $c_1(x)$ and $c_2(x)$, respectively. Then the code $C_k=\operatorname{Row}[G_{T_g^n}\mid D_kG_{T_g^n}J_n]$ is an $(n,c_1|n,c_2)-$right polycyclic code with respect to associated vectors $\mathbf{c}_1$ and $\mathbf{c}_2$ for every $k=0,1,\ldots,m-1$, where $\operatorname{Row}[G_{T_g^n}\mid D_kG_{T_g^n}J_n]$ denotes the linear code generated by $[G_{T_g^n}\mid D_kG_{T_g^n}J_n]$. Moreover, $D_k$ is involutory, that is, $D_k^2=I_m$.
	\end{propo}
	\begin{proof}
		As $n=s+m$, we have $x^n=x^m x^s$. So, $x^n-g(x)(x^m+1)=x^m(x^s-g(x))-g(x)$. Since $g(x)$ is monic of degree $s,$ the  polynomial $c_1(x)$ has degree strictly less than $n$. Moreover, from the definition of $c_1(x)$, we have $x^n-c_1(x)=g(x)(x^m+1)$. This imples that $g(x)$ divides $x^n-c_1(x)$. Similarly, $g^*(x)$ divides $x^n-c_2(x)$, because $x^n-c_2(x)=g^*(x)(x^m+1)$. Thus multiplication by $x$ modulo the corresponding polycyclic relation gives $x^mg(x)=-g(x)$ and $x^mg^*(x)=-g^*(x)$. Consequently, under the $\mathbf{c_1}$-polycyclic shifts, the sequence of generators $g(x),xg(x),\ldots,x^{m-1}g(x)$ is transformed cyclically as $g(x)\mapsto xg(x)\mapsto\cdots\mapsto x^{m-1}g(x)\mapsto-g(x)$. Therefore, the corresponding coefficient matrix satisfies $\tau_{\mathbf{c_1}}(G_{T_g^n})=S_mG_{T_g^n}$. Hence, by Lemma \ref{polycyclic shift in terms of Sm} for every $x\in R^m$, we have $\tau_{\mathbf{c_1}}(xG_{T_g^n})=xS_mG_{T_g^n}=(xS_m)G_{T_g^n}$. Similarly, from $x^mg^*(x)=-g^*(x)$ we obtain $\tau_{\mathbf{c_2}}(G_{{T_{g^*}}^n})=S_mG_{{T_{g^*}}^n}$ and consequently $\tau_{\mathbf{c_2}}(xG_{{T_{g^*}}^n})=(xS_m)G_{{T_{g^*}}^n}$.
		Next, by Lemma \ref{MainLemma} with $D=J_m$ the relation between the generator matrices associated with a polynomial and its reciprocal gives $G_{{T_{g^*}}^n}=J_mG_{T_g^n}J_n$. Hence $D_kG_{T_g^n}J_n=S_m^kJ_mG_{T_g^n}J_n=S_m^kG_{{T_{g^*}}^n}$. Thus the generator matrix of $C_k$ can be written as $G_k=[G_{T_g^n}\mid S_m^kG_{{T_{g^*}}^n}]$. Let $(u\mid v)\in C_k$. Then there exists $x\in R^m$ such that $(u\mid v)=xG_k=(xG_{T_g^n}\mid xS_m^kG_{{T_{g^*}}^n})$. The first component transforms under $\tau_{\mathbf{c_1}}$ as $\tau_{\mathbf{c_1}}(u)=xS_mG_{T_g^n}=(xS_m)G_{T_g^n}$. For the second component, we obtain $\tau_{\mathbf{c_2}}(v)=\tau_{\mathbf{c_2}}(xS_m^kG_{{T_{g^*}}^n})=xS_m^kS_mG_{{T_{g^*}}^n}$. Since $S_m^k$ and $S_m$ are powers of the same matrix, they commute, and therefore $xS_m^kS_mG_{{T_{g^*}}^n}=xS_mS_m^kG_{{T_{g^*}}^n}=(xS_m)S_m^kG_{{T_{g^*}}^n}$. Consequently, $(\tau_{\mathbf{c_1}}(u)\mid\tau_{\mathbf{c_2}}(v))=(xS_mG_{T_g^n}\mid(xS_m)S_m^kG_{{T_{g^*}}^n})=(xS_m)[G_{T_g^n}\mid S_m^kG_{{T_{g^*}}^n}]=(xS_m)G_k\in C_k$, where $G_k= [G_{T_g^n}\mid D_kG_{T_g^n}J_n]$. Thus $C_k$ is an $(n,c_1|n,c_2)-$right polycyclic code. 
		Finally, since $D_k=S_m^kJ_m$, it follows by Lemma \ref{Sm lemma} that the reversal matrix satisfies $J_mS_mJ_m=S_m^{-1}$. We have $J_mS_m^kJ_m=S_m^{-k}$. Therefore $D_k^2=S_m^kJ_mS_m^kJ_m=S_m^kS_m^{-k}=I_m$. Hence each $D_k$ is involutory.
	\end{proof}
	\begin{thm}\label{poly main construction}Let $R$ be a finite commutative ring with unity and $g(x)=1+z_1x+\cdots+z_{s-1}x^{s-1}+x^s\in R[x]$, be a monic polynomial of degree $s<n.$ Let $n=s+m$ with $m\geq 1$. Let $S_m$ be the $m\times m$ matrix defined by $S_m=\begin{bmatrix}0&1&0&\cdots&0\\
			0&0&1&\cdots&0\\ 
			\vdots&\vdots&\ddots&\vdots&1\\
			-1&0&\cdots&0&0
		\end{bmatrix}.$
		For $k=0,1,\ldots,m-1$, let $D_k=S_m^kJ_m$. Let $c_1(x)=x^n-g(x)(x^m+1)$ and $c_2(x)=x^n-g^*(x)(x^m+1)$ and $\mathbf{c}_1$ and $\mathbf{c}_2$ be the vectors whose entries are the coefficients of $c_1(x)$ and $c_2(x)$, respectively. Then the code $C_k=\operatorname{Row}[G_{T_g^n}\mid D_kG_{T_g^n}J_n]$ is a linear reversible $(n,c_1|n,c_2)-$right polycyclic code with respect to associated vectors $\mathbf{c}_1$ and $\mathbf{c}_2$  for every $k=0,1,\ldots,m-1$.
	\end{thm}
	\begin{proof} If we take $G_{T_g^n}=M$ and $D=D_k$ in Lemma \ref{linear2}, then by definition of the $D-$reflected matrix, $D_kG_{T_g^n}J_n$ is the $D-$ reflected matrix of $M$, that is, $D_kG_{T_g^n}J_n=M_D'.$ Hence, $C_k=\operatorname{Row}[G_{T_g^n}\mid D_kG_{T_g^n}J_n]$ is a reversible linear code for every $k=0,1,\ldots,m-1$. Further, by Proposition \ref{partwise polycylic}, $C_k$ is a $(n,c_1|n,c_2)-$right polycyclic code for every $k=0,1,\ldots,m-1$. It follows that $C_k=\operatorname{Row}[G_{T_g^n}\mid D_kG_{T_g^n}J_n]$ is a linear reversible $(n,c_1|n,c_2)-$right polycyclic code for every $k=0,1,\ldots,m-1$.
		
	\end{proof}
	\begin{exl}\label{main example poly} In the following, we write $16$ MDS out of $243$ MDS $(12,c_1|12,c_2)-$ right polycyclic codes using Sage Math \cite{Sage} computations with the restriction of only 3000 polynomials over the field $F_4$ for computational feasibility. 
		
		\begin{table}[htbp]
			\centering
		\small
			\caption{MDS $(12,c_1|12,c_2)-$right polycyclic reversible codes}
			$\renewcommand{\arraystretch}{1.3}       
			\begin{array}{|c|c|c|c|c|c|c|c|c|c|}
				\hline
				
				\textbf{No} & \textbf{q} & \mathbf{g(x)} & \mathbf{s} & \mathbf{m} & \mathbf{j} & \mathbf{n} & \mathbf{k} & \mathbf{d} & \mathbf{MDS}\\
				\hline
				1 & 4 & x^5+ax^4+ax^3+ax^2+ax+a & 5 & 1 & 0 & 12 & 1 & 12 & \mathrm{Yes}\\ 
				2 & 4 & x^5+(a+1)x^4+ax^3+ax^2+ax+a & 5 & 1 & 0 & 12 & 1 & 12 & \mathrm{Yes}\\
				3 & 4 & x^5+x^4+ax^3+ax^2+ax+a & 5 & 1 & 0 & 12 & 1 & 12 & \mathrm{Yes}\\
				4 & 4 & x^5+ax^4+(a+1)x^3+ax^2+ax+a & 5 & 1 & 0 & 12 & 1 & 12 & \mathrm{Yes}\\
				5 & 4 & x^5+(a+1)x^4+(a+1)x^3+ax^2+ax+a & 5 & 1 & 0 & 12 & 1 & 12 & \mathrm{Yes}\\
				6 & 4 & x^5+x^4+(a+1)x^3+ax^2+ax+a & 5 & 1 & 0 & 12 & 1 & 12 & \mathrm{Yes}\\
				7 & 4 & x^5+ax^4+x^3+ax^2+ax+a & 5 & 1 & 0 & 12 & 1 & 12 & \mathrm{Yes}\\
				8 & 4 & x^5+(a+1)x^4+x^3+ax^2+ax+a & 5 & 1 & 0 & 12 & 1 & 12 & \mathrm{Yes}\\
				9 & 4 & x^5+x^4+x^3+ax^2+ax+a & 5 & 1 & 0 & 12 & 1 & 12 & \mathrm{Yes}\\
				10 & 4 & x^5+ax^4+ax^3+(a+1)x^2+ax+a & 5 & 1 & 0 & 12 & 1 & 12 & \mathrm{Yes}\\
				11 & 4 & x^5+(a+1)x^4+ax^3+(a+1)x^2+ax+a & 5 & 1 & 0 & 12 & 1 & 12 & \mathrm{Yes}\\
				12 & 4 & x^5+x^4+ax^3+(a+1)x^2+ax+a & 5 & 1 & 0 & 12 & 1 & 12 & \mathrm{Yes}\\
				13 & 4 & x^5+ax^4+(a+1)x^3+(a+1)x^2+ax+a & 5 & 1 & 0 & 12 & 1 & 12 & \mathrm{Yes}\\
				14 & 4 & x^5+(a+1)x^4+(a+1)x^3+(a+1)x^2+ax+a & 5 & 1 & 0 & 12 & 1 & 12 & \mathrm{Yes}\\
				15 & 4 & x^5+x^4+(a+1)x^3+(a+1)x^2+ax+a & 5 & 1 & 0 & 12 & 1 & 12 & \mathrm{Yes}\\
				16 & 4 & x^5+ax^4+x^3+(a+1)x^2+ax+a & 5 & 1 & 0 & 12 & 1 & 12 & \mathrm{Yes}\\ \hline
			\end{array}
			$  
		\end{table}    		      
		
	\end{exl}
	
	\section{Conclusion and Future Scope}    	  
	In this work, we have presented $T_{D,g}^n-$set construction technique for reversible linear codes that generalizes the $T^n-$ set construction given in \cite{Oztas2}. The generalization is then extended to the construction of reversible DNA codes. The generalized construction technique broadens the range of available constructions and allows to obtain improved code parameters, including maximum-distance-separable (MDS) codes in suitable cases. Moreover, we have obtained a characterization of polynomial $g$ for the $T_{D,g}^n-$set construction to be a self-dual linear reversible code. The concatenation of matrices approach in \cite{Chaturvedi}, has been extended to the construction of linear complementary dual (LCD) reversible linear codes. Further, we have introduced the notion of $(n_1,c_1|n_2,c_2)-$right polycyclic codes and given a construction technique based on $T_{D,g}^n-$set, which also incorporates reversibility. Overall, these constructions demonstrate how suitable concatenation techniques can be employed to impose structural constraints like reversibility, self duality, LCD property while retaining good coding parameters.
	
	The importance of these constructions extends beyond the existence of codes with prescribed algebraic properties. Reversible codes have applications in coding schemes where a codeword and its reversal must both belong to the code. DNA codes additionally require compatibility with biological constraints like GC balanced constraint. Such codes have potential applications in DNA-based data storage and DNA computational techniques. The generalized DNA constructions considered here provide a wider framework for investigating these requirements alongside reversibility and distance properties.
	
	The constructions of self-dual and LCD reversible codes are also of interest in cryptography and secure information processing. LCD codes have been investigated in connection with code-based cryptographic systems, side-channel-attack and fault injectioin attack resistant implementations, and related security-oriented applications. Self-dual codes, in turn, are relevant to the algebraic constructions of quantum error correction and other coding-theoretic settings involving duality constraints. The constructions developed in this work may therefore provide a useful base for exploring codes that combine reversibility with these additional properties.
	
	Several directions for further study arise naturally from the present work, as follows.
	
	(1). The parameter ranges of the $T_{D,g}^n$-set construction may be examined systematically to determine sufficient conditions for obtaining MDS, almost-MDS, or other near-optimal codes.
	
	(2). Characterization of the choices of polynomials, matrices, and underlying finite rings or fields that yield the best minimum distances for prescribed lengths and dimensions.
	
	(3). The conditions under which the $T_{D,g}^n$-set construction for DNA codes meet other  DNA code properties like reverse complement and GC-balanced constraint.
	
	(4). The $(n_1,c_1|n_2,c_2)-$right polycyclic code framework may be extended to more than two blocks over an arbitrary finite commutative ring with unity, allowing different polycyclic shifts to act on different components. This raises the question of identifying general conditions under which such multi-block constructions preserve reversibility, duality properties, and desirable distance bounds.
	
	\section*{Acknowledgement} The second author is  grateful to the Council of Scientific and Industrial Research, India, for awarding the Junior Research Fellowship, file number: 09/0001 (20883)/2025-EMR-I.


\begin{thebibliography}{99}
		\bibitem{Abolution}Aboluion, N., Smith, D.H. and Perkins, S., 2012. Linear and nonlinear constructions of DNA codes with Hamming distance d, constant GC-content and a reverse-complement constraint. Discrete Mathematics, 312(5), pp.1062-1075.
		\bibitem{Adleman}Adleman, L.M., 1994. Molecular computation of solutions to combinatorial problems. science, 266(5187), pp.1021-1024.
		\bibitem{AdlemanDES}Adleman, L.M., Rothemund, P.W., Roweis, S. and Winfree, E., 1999. On applying molecular computation to the data encryption standard. Journal of computational biology, 6(1), pp.53-63.
		\bibitem{Aydin}Aydin, N., Liu, P. and Yoshino, B., 2022. Polycyclic codes associated with trinomials: good codes and open questions: N. Aydin et al. Designs, Codes and Cryptography, 90(5), pp.1241-1269.
		
		\bibitem{Boneh}Boneh, D., Dunworth, C. and Lipton, R.J., 1995. Breaking DES using a molecular computer. DNA based computers, 27, pp.37-66.
		\bibitem{Brawley}Brawley, J.V. and Gamble, R.O., 1978. Involutory matrices over finite commutative rings. Linear Algebra and its Applications, 21(2), pp.175-188.
		
		\bibitem{Chaturvedi}Chaturvedi, A.K. and Pandey, S., 2026. On concatenation of matrices for reversible linear codes over a finite commutative ring and applications to DNA codes. arXiv preprint arXiv:2606.10633.	
		\bibitem{De Broer}De Boer, M.A., 1996. Almost MDS codes. Designs, Codes and Cryptography, 9(2), pp.143-155.
		\bibitem{Fotue}Fotue-Tabue, A., Martínez-Moro, E. and Blackford, J.T., 2020. On polycyclic codes over a finite chain ring. Advances in Mathematics of Communications, 14(3), pp.455-466.
		
		\bibitem{Frutos}Frutos, A.G., Liu, Q., Thiel, A.J., Sanner, A.M.W., Condon, A.E., Smith, L.M. and Corn, R.M., 1997. Demonstration of a word design strategy for DNA computing on surfaces. Nucleic Acids Research, 25(23), pp.4748-4757.
		
		\bibitem{Gaborit}Gaborit, P. and King, O.D., 2005. Linear constructions for DNA codes. Theoretical Computer Science, 334(1-3), pp.99-113.
		\bibitem{Golub}Golub, G.H. and Van Loan, C.F.. Matrix computations. JHU press, 2013.
		\bibitem{King}King, O.D., 2003. Bounds for DNA codes with constant GC-content. arXiv preprint math/0306197.
		\bibitem{Xing} Ling, S. and Xing, C. Coding theory: a first course. Cambridge university press, 2004.
		\bibitem{Mansuripur}Mansuripur, M., Khulbe, P.K., Kuebler, S.M., Perry, J.W., Giridhar, M.S. and Peyghambarian, N., 2003, May. Information storage and retrieval using macromolecules as storage media. In Optical data storage (p. TuC2). Optica Publishing Group.
		\bibitem{Massey}Massey, J.L., 1992. Linear codes with complementary duals. Discrete Mathematics, 106, pp.337-342.
		\bibitem{matsuoka}MATSUOKA, M., 2012. Polycyclic codes and sequential codes, Proceedings of the 44th Symposium on Ring Theory and Representation Theory, September 25(Sun.)- 27(Tue.), 2011
		Okayama University, Japan.
		\bibitem{Oztas}Oztas, E.S., Alali, A.S., Ali, S., Azeem, M., Asri, M.S. and Wong, K.B., 2025. A Matrix-Theoretic Approach to the Construction of Reversible DNA Codes Over $F_{4^{2t}}$. IEEE Access, 13, pp.209059-209071.
		\bibitem{Oztas2}Oztas, E.S., Alali, A.S., Ali, S., Azeem, M., Asri, M.S. and Wong, K.B., 2025. On the construction of reversible DNA codes over $F_{ 4 ^{2 m}}$ via $T^n$-set codes. AIMS Mathematics, 10(12), pp.29424-29453.
		\bibitem{Siap}Oztas, E.S. and Siap, I., 2013. Lifted Polynomials Over $\mathbb{F}_{16}$ and Their Applications to DNA Codes. Filomat, 27(3), pp.459-466.
		
		\bibitem{Saif}Saif, S.H. and Aldossari, S., 2026. Optimal Generalized Quasi-Polycyclic Codes over $F_q+uF_q$. Mathematics, 14(5), p.816.
		\bibitem{Shi M.}Shi, M., Li, X., Sepasdar, Z. and Solé, P., 2020. Polycyclic codes as invariant subspaces. Finite Fields and Their Applications, 68, p.101760.
		\bibitem{Shi}Shi, M., Xu, L. and Solé, P., 2020. Construction of isodual codes from polycirculant matrices: M. Shi et al. Designs, Codes and Cryptography, 88(12), pp.2547-2560.
		
		\bibitem{Smith}Smith, D.H., Aboluion, N., Montemanni, R. and Perkins, S., 2011. Linear and nonlinear constructions of DNA codes with Hamming distance d and constant GC-content. Discrete mathematics, 311(13), pp.1207-1219.
		\bibitem{Song}Song, W., Cai, K., Zhang, M. and Yuen, C., 2018. Codes with run-length and GC-content constraints for DNA-based data storage. IEEE Communications Letters, 22(10), pp.2004-2007.
		\bibitem{Suxsena}Suxena, K., Prakash, O., Debnath, I. and Solé, P., 2025. Optimal GQPC codes over the finite field Fq. Mathematics, 13(22), p.3655.
		\bibitem{Sage}The Sage Developers, SageMath, the Sage Mathematics Software System (Version 9.3).	
		\bibitem{Wu}Wu, R., Shi, M. and Solé, P., 2022. On the structure of 1-generator quasi-polycyclic codes over finite chain rings. Journal of Applied Mathematics and Computing, 68(5), pp.3491-3503.
		
	\end{thebibliography}
\end{document}